\documentclass[acmsmall]{acmart}
\usepackage{wrapfig}
\usepackage[caption=false]{subfig}
\usepackage[ruled,vlined]{algorithm2e}
\usepackage{multirow}
\usepackage{url}

\AtBeginDocument{%
  \providecommand\BibTeX{{%
    \normalfont B\kern-0.5em{\scshape i\kern-0.25em b}\kern-0.8em\TeX}}}

\setcopyright{acmcopyright}
\copyrightyear{2020}
\acmYear{2020}
\acmDOI{10.1145/1122445.1122456}

\begin{document}

\title{Mez: A Messaging System for Latency-Sensitive Multi-Camera Machine Vision at the IoT Edge}

\author{Anjus George}
\email{ageorg28@uncc.edu}
\author{Arun Ravindran}
\email{Arun.Ravindran@uncc.edu}
\author{Mat\'ias Mendieta}
\email{mmendiet@uncc.edu}
\author{Hamed Tabkhi}
\email{htabkhiv@uncc.edu}
\affiliation{%
  \institution{University of North Carolina at Charlotte}
  \streetaddress{9201 University City Blvd}
  \city{Charlotte}
  \state{North Carolina}
  \postcode{28223}
}

\renewcommand{\shortauthors}{George, et al.}

\begin{abstract}
Mez is a publish-subscribe messaging system for latency sensitive multi-camera machine vision at the IoT Edge. Unlike existing messaging systems, Mez allows applications to specify latency, and application accuracy bounds. Mez implements a network latency controller that dynamically adjusts the video frame quality to satisfy latency, and application accuracy requirements. Additionally, the design of Mez utilizes application domain specific features to provide low latency operations. Experimental evaluation on an IoT Edge testbed with a pedestrian detection machine vision application indicates that Mez is able to tolerate latency variations of up to 10x with a worst-case reduction of 4.2\% in the application accuracy F1 score metric.

\end{abstract}



\maketitle

\section{Introduction}
\label{sec:introduction}

The recent emergence of powerful machine vision algorithms based on Deep Learning has made possible Internet-of-Things (IoT) applications that utilize machine vision for a variety of challenging tasks including autonomous driving, pedestrian safety, public security, and occupational health and safety. Such applications involve computationally intensive processing of streaming videos from cameras operating 24x7x365. Assuming a modest frame rate of 5 fps, and a 500 kB frame size, 216 GB of data is generated per camera per day (19.5 Mbps per camera). Often, multiple cameras are needed to provide adequate area coverage for subject tracking and overcoming occlusions \cite{multi_camera_review, multi_camera_tacking}. Additionally, the aforementioned applications tend to be latency sensitive - that is, the processing of video frames needs to be done within a short time window for the results to be useful. The duration of the time window depends on the speed of the event (for example, tracking a high speed vehicle vs. tracking a pedestrian), and the response time needed for useful actions (for example, sounding an alert before vs. after event). Furthermore, while many of these applications enable broader smart city initiatives, and social good, significant privacy concerns exist regarding the potential misuse of the collected video data \cite{Zhang17}. 

Despite the considerable computing power available in the Cloud, the use of Cloud computing for IoT machine vision applications is hindered by the high network latency to access a remote data center (typically hundreds of milliseconds)\cite{lat_cog_assistance}, and constraints in the upload bandwidth (typically tens of Mbps). Moreover, privacy and legal concerns place limitations on sharing of sensitive data to remote servers controlled by external entities. IoT machine vision applications are thus an ideal candidate for the Edge computing paradigm \cite{case_for_VM, swarm_at_edge_cloud, edge_comp_challenge, Fog-edge, Fog-computing} , where most of the processing of the video streams happen in the vicinity of the camera. The Cloud may still play a role in aggregating detected events from multiple IoT Edge deployments, both for performing batch analytics, and for archival purposes. The localized processing of video streams at the Edge potentially allows for low latency operation, overcomes bandwidth limitations by reducing the size of the data that needs to be sent to the Cloud, and helps addresses privacy and legal concerns by eliminating the need to share raw video frames with Cloud vendors.

\begin{wrapfigure}{r}{0.5\textwidth}
  \begin{center}
    \includegraphics[width=0.35\textwidth]{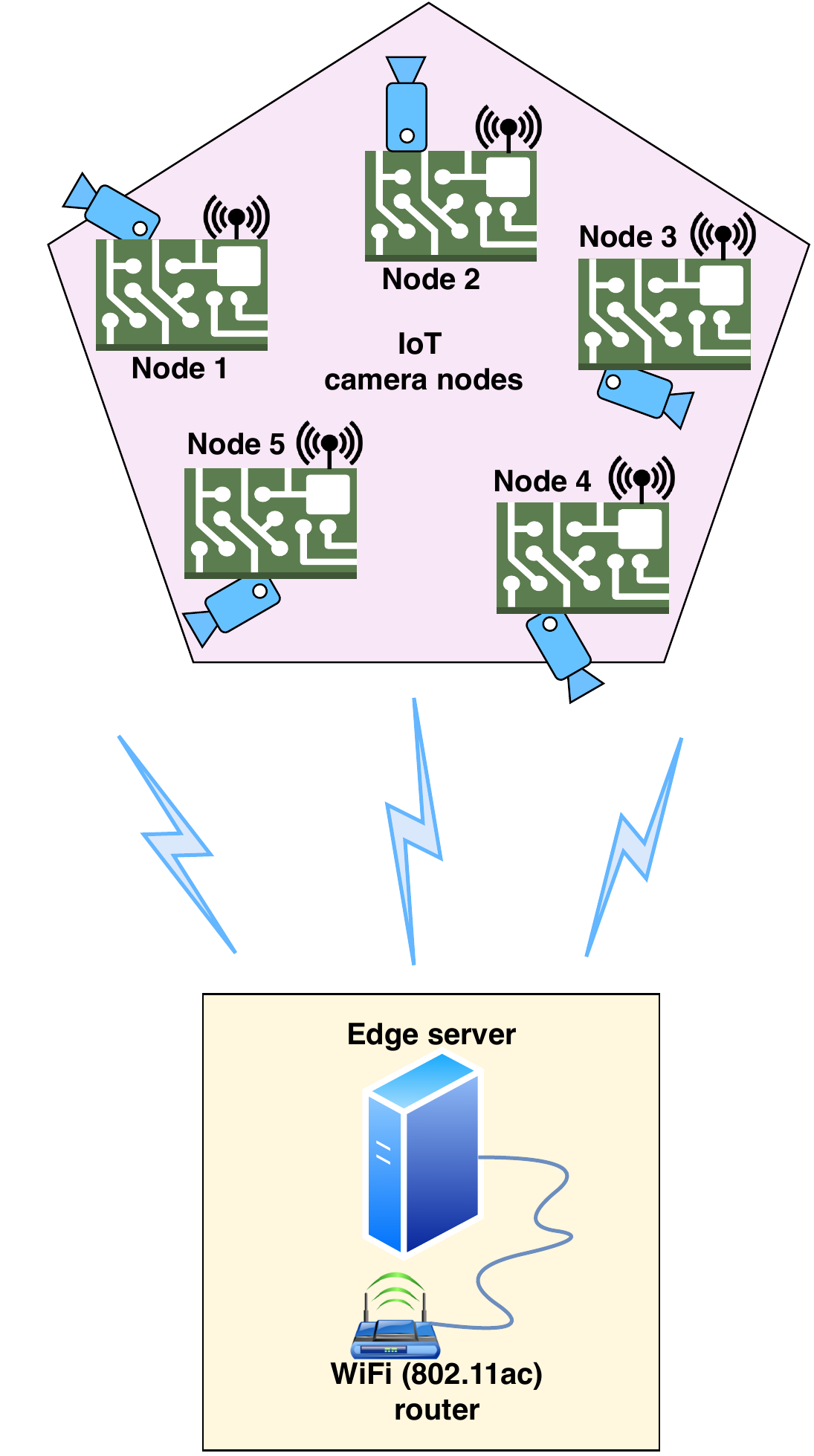}
  \end{center}
  \caption[Distributed machine vision at the Edge.]{\label{fig:physical_arch}System architecture for machine vision at the IoT Edge. IoT nodes equipped with cameras record and transmit video frames to an Edge server through a WiFi (802.11ac) wireless router. The Edge server runs  machine vision algorithms on the received video frames for object detection, tracking and event prediction.}
\end{wrapfigure}

\begin{figure}
    \center{\includegraphics[width=0.7\textwidth]{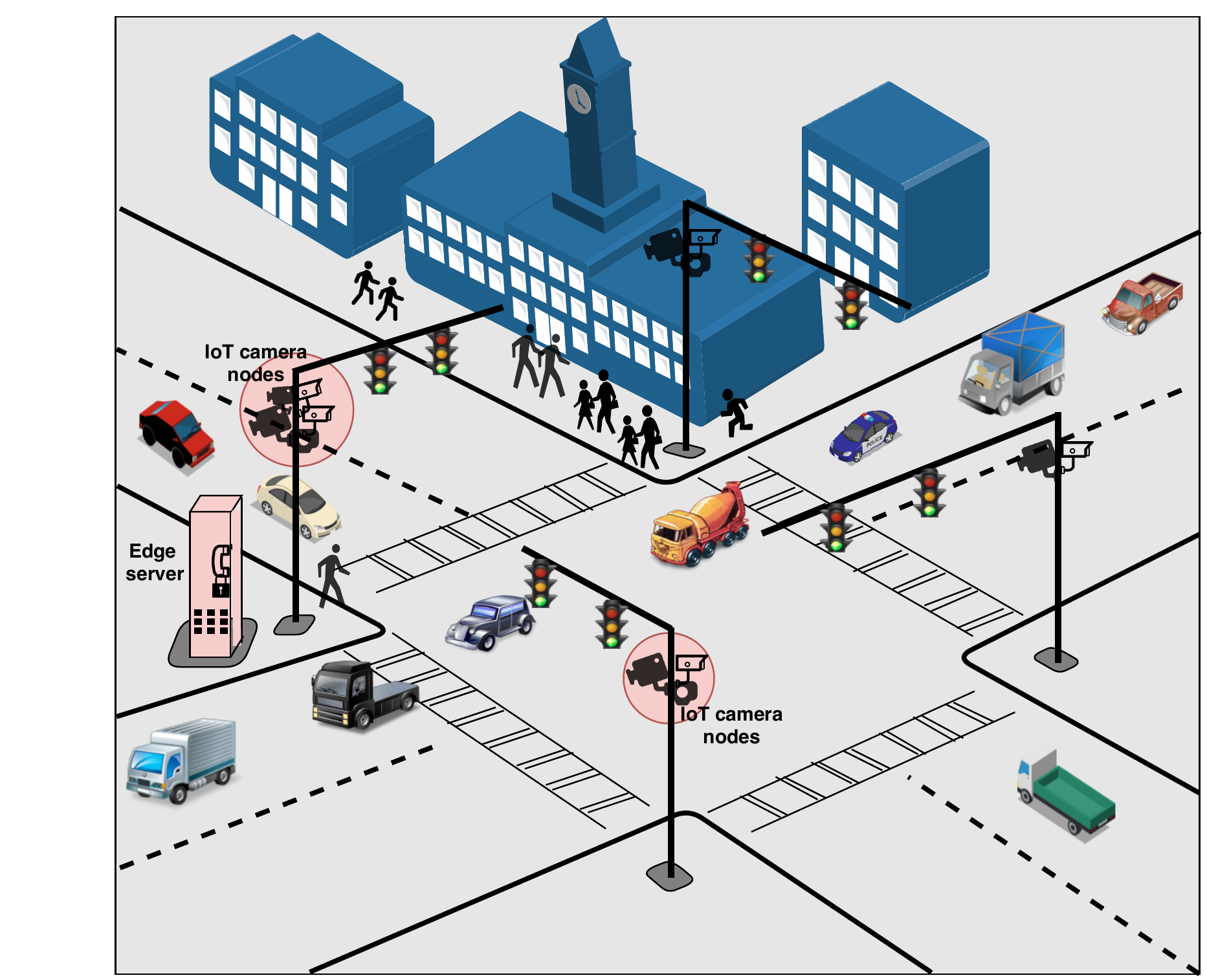}}
    \caption{\label{fig:traffic_intersection}Multi-camera machine vision at the IoT Edge for pedestrian safety.}
\end{figure}

Figure \ref{fig:physical_arch} shows the system architecture for machine vision at the IoT Edge. Low power IoT nodes equipped with cameras, stream live videos of the area under observation to an Edge server equipped with GPUs through a wireless network. The Edge server aggregates the individual video streams from multiple cameras, and run machine vision applications for object detection, tracking, and event prediction. For an application such as pedestrian safety (see Figure \ref{fig:traffic_intersection}), the cameras are mounted on traffic signal posts at street intersections, and the Edge server is housed in a traffic signal box.

The Edge differs from the Cloud in some significant ways - unlike high speed wired networks in data centers, wireless networks at the Edge allow flexible installation of cameras at a lower cost. Also, cost reasons motivate the use of wireless technologies such as WiFi (802.11ac) and Bluetooth (BLE) that operate in the free unlicensed bands. In contrast 4G and 5G wireless technologies operate in the licensed bands, and requires paid subscriptions to cellular network vendors. Additionally, space, cost, and power supply constraints limit the hardware redundancy available at the Edge. Moreover, ensuring the physical security of the hardware is potentially challenging at the Edge due to deployments in unsecured environments.

In this paper, we explore the characteristics of an IoT Edge middleware layer that provides a suitable abstraction for machine vision application developers to deploy vision applications that consume video streams from one or more cameras. Since the applications are latency sensitive, the middleware layer should provide a means for applications to specify the latency requirements. The middleware then makes the best-effort to guarantee the specified latency. The use of wireless technologies such as WiFi makes this particularly challenging, due to the large latency variations in the wireless channel. We propose the use of publish-subscribe (pub-sub) messaging system with storage, as a candidate IoT Edge middleware. A pub-sub system decouples publishers (cameras) from subscribers (machine vision applications). IoT camera nodes publish video frames to topics identified by a camera ID. Applications subscribe to one or more topics as needed. The storage layer allows temporal decoupling of publishers and subscribers, allowing subscriber applications to access past video frames. In Cloud computing, such pub-sub systems are widely deployed to handle real-time data feeds, and as message brokers between microservices. Open source examples of such messaging systems include Kafka \cite{KrepsKafka}, NATS \cite{Nats}, and RabbitMQ \cite{RabbitMQ}. Kafka is designed for high throughput, NATS  targets low latency, and RabbitMQ allows for complex routing between publishers and subscribers. However, the existing messaging systems are built for the Cloud, where machines communicate over low latency wired networks (Gigabit Ethernet, Infiniband), and as such do not provide mechanisms to guarantee latency when operating in wireless channels with large latency variations.

We introduce Mez, a pub-sub messaging system specifically designed for machine vision applications at the IoT Edge. The key design themes of Mez are -
\begin{itemize}
    \item Approximate computing - Mez exploits the trade off between video frame transfer latency from the IoT camera node to the Edge server, and video frame quality (approximate computing) inherent in machine vision applications. A lower quality video frame has a smaller size, and hence can be transferred with lower network latency. If a lower quality frame can provide acceptable accuracy, then a lower quality frame could be transferred from the IoT camera node to the Edge server during conditions of high channel interference. It should be noted that the application developer has to determine acceptable latency-accuracy tradeoffs. Mez provides a means for applications to specify the upper bound on the latency, and the lower bound on the accuracy requirements.  The accuracy in turn translates to the quality of the video frame that Mez has to deliver.
    \item Adaptive computing - Mez constantly monitors the operating conditions of the wireless channel. When channel interference is high, Mez automatically adapts the  quality of the video frames such that both the frame transfer latency and accuracy specifications are met.
    \item Domain specific design - Mez  employs an on-demand video frame transfer from the IoT camera node to the Edge server to minimize wireless channel interference. Furthermore, Mez uses an in-memory log based storage that exploits application domain specific characteristics to implement a simple low latency storage.
\end{itemize}

We have implemented Mez on an IoT Edge testbed with multiple IoT camera nodes, and a single GPU equipped Edge server. The source code is available from our GitHub repository\footnote{\url{https://github.com/Ann-Geo/Mez}}. Experimental results indicate that for pedestrian detection application with the OpenPose  multi-person 2D pose detection benchmark \cite{openpose1, openpose2, openpose3, openpose4}, Mez is able to achieve the target latency in the presence of up to 10x increase in channel interference with an application accuracy degradation of at most 4.2\%. In contrast, the state-of-the-art NATS messaging framework suffers from latency degradation as the number of IoT camera nodes scale.

The rest of this paper is organized as follows - In Section \ref{sec:chara_of_latency} we experimentally characterize the latency issues on an IoT Edge testbed for vision applications. Section \ref{sec:mez_architecture} presents the Mez API and architecture. Section \ref{sec:mez_design} describes the detailed design of Mez including the adaptive latency controller, and the low latency storage layer. Section \ref{sec:results} presents the experimental results and evaluation of Mez on the IoT Edge testbed. In Section \ref{sec:discussion} we discuss the different design decisions made in Mez, and explore alternative approaches. Section \ref{sec:related_work} provides a brief review of related work in Edge computing, approximate computing, and distributed messaging systems. Section \ref{sec:conclusion} concludes the paper.


\section{Characterization of Wi-Fi latency at the Edge}
\label{sec:chara_of_latency}

In this section, we present the study of the impact on the latency of video frames transferred over WiFi from IoT camera nodes to the Edge server (referred to henceforth as network latency for brevity) due to multiple factors - (1)  interference by peer IoT camera nodes, (2) video scene dynamics, (3) video frame rate, and (4) video frame quality.

\subsection{IoT Edge Test Bed}
\label{sec:testbed}

 We set up an Edge testbed similar to the IoT Edge machine vision system shown in Figure \ref{fig:physical_arch}. Our Edge test bed consists of five IoT camera nodes equipped with 8-core ARMv8.2 based embedded Nvidia Jetson AGX Xavier \cite{xavier} boards, and an Edge server. A workstation equipped with an Nvidia Titan V GPU serves as the Edge server. The embedded boards and the workstation run Linux. The wireless link consists of a NETGEAR Nighthawk XR700 access point that uses 802.11ac (5 GHz) WiFi standard. The Edge server is connected to the access point through Ethernet, while the IoT camera nodes connect to the access point through the 802.11ac WiFi link. The IoT camera nodes are placed at 6m from the access point.

We use two publicly available video datasets - JAAD \cite{JAAD} and DukeMTMC \cite{dukemtmc} for latency characterization. The JAAD dataset consists of videos of pedestrian movement in public spaces captured under various camera types and qualities in different weather/lighting conditions. The DukeMTMC data set consists of 1080p videos recorded at 60 fps from 8 static cameras deployed on the Duke University campus. To perform the Edge latency measurements, we chose video clips with three different scene dynamics - simple, medium, and complex, from both JAAD and DukeMTMC datasets.

\begin{figure}
  \begin{center}
    \includegraphics[width=0.6\textwidth]{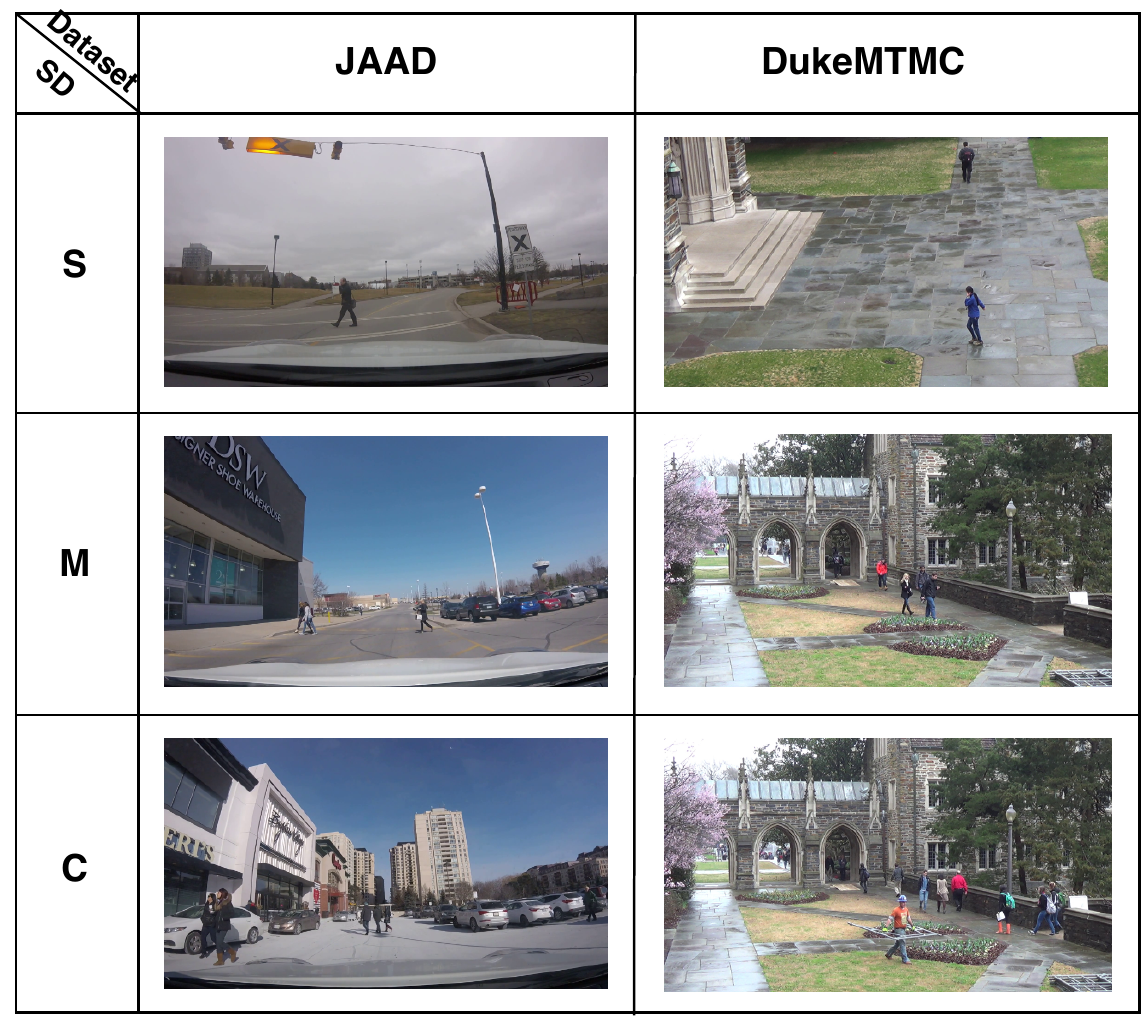}
  \end{center}
  \caption{\label{fig:original_images}Sample images from JAAD and DukeMTMC dataset with simple (S), medium (M) and complex (C) scene dynamics (SD) showing pedestrians at public spaces such as traffic intersections, parking lots, and public buildings. }
\end{figure}

In order to cluster the video frames in the two data sets as simple, medium and complex, a k-means clustering approach was implemented. For DukeMTMC, all frames from two of the cameras (cameras 5 and 6) were k-means clustered using the mean and standard deviation of the bounding box areas in a scene. A frame sequence of 100 frames was then randomly selected from each cluster for network latency measurements. A similar approach was followed for the JAAD dataset. Figure \ref{fig:original_images} shows a representational sample of images from JAAD and DukeMTMC datasets.

A Golang gRPC \cite{grpc} based client and multi-threaded server was deployed at the IoT camera node and the Edge server respectively to facilitate video frame transfer, and perform network latency measurements. The wireless network latency of video frame transfer is measured by sending timestamped images from IoT camera node to the Edge server. The latency is calculated as time difference $t_{Received} - t_{Send}$. The IoT camera nodes are time synchronized to the Edge server before starting the network latency measurements using the PTP network level time synchronization protocol capable of microsecond accuracy \cite{ptp}.

\subsection{Evaluating the impact of peer IoT nodes, scene dynamics, and video frame rate on network latency}
\label{sec:interference}

In the measurements described below, we measure the network latency experienced by IoT camera node 1 (see Figure \ref{fig:physical_arch}) due to the 4 peer IoT camera nodes. All latency measurements are at the 95th percentile with video frames transmitted at 5 fps.

\begin{figure}
\centering
\subfloat[]{\includegraphics[width=0.48\textwidth, keepaspectratio
]{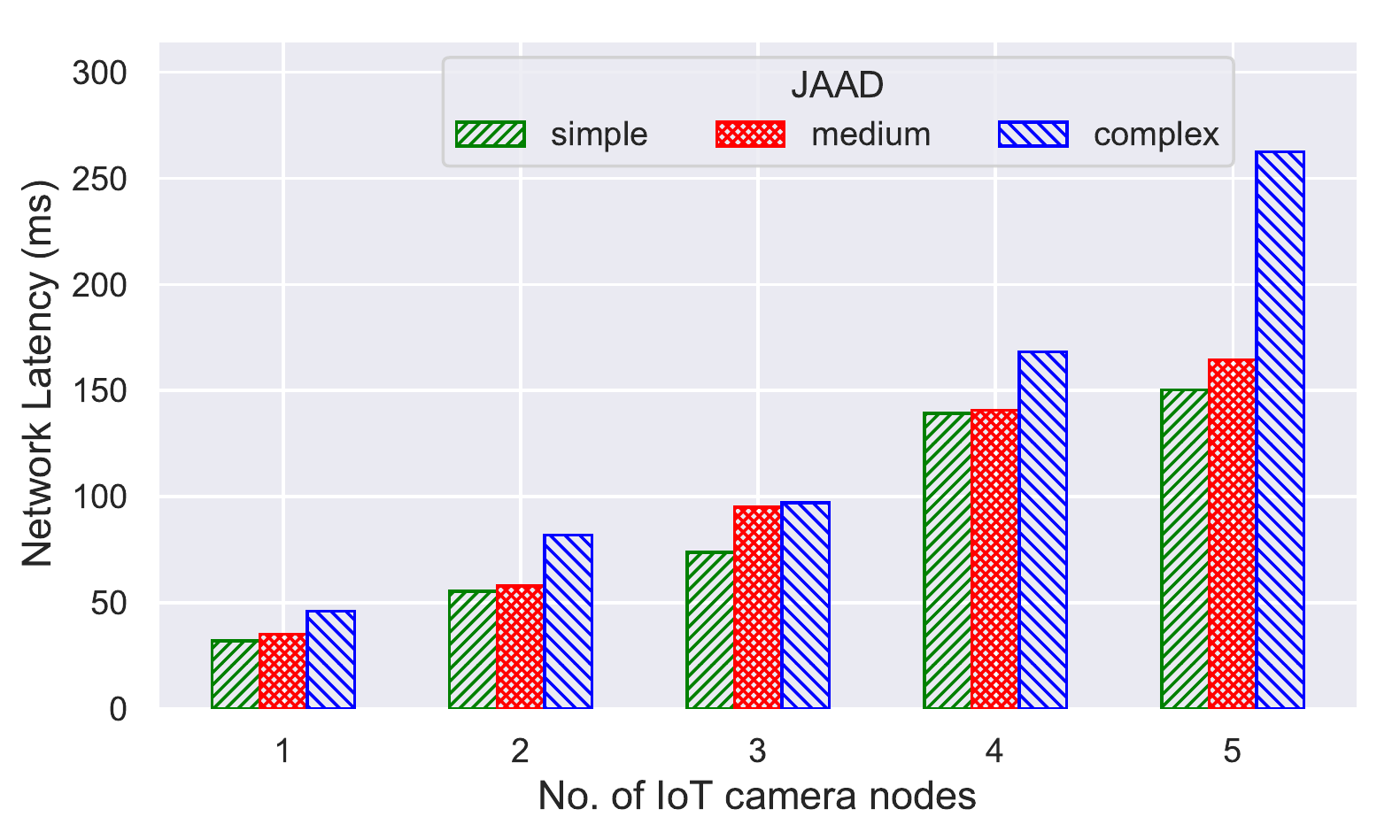}\label{fig:lat_vs_num_nodes_scene_dyn_jaad}}\hspace{2pt}
\subfloat[]{\includegraphics[width=0.48\textwidth, keepaspectratio]{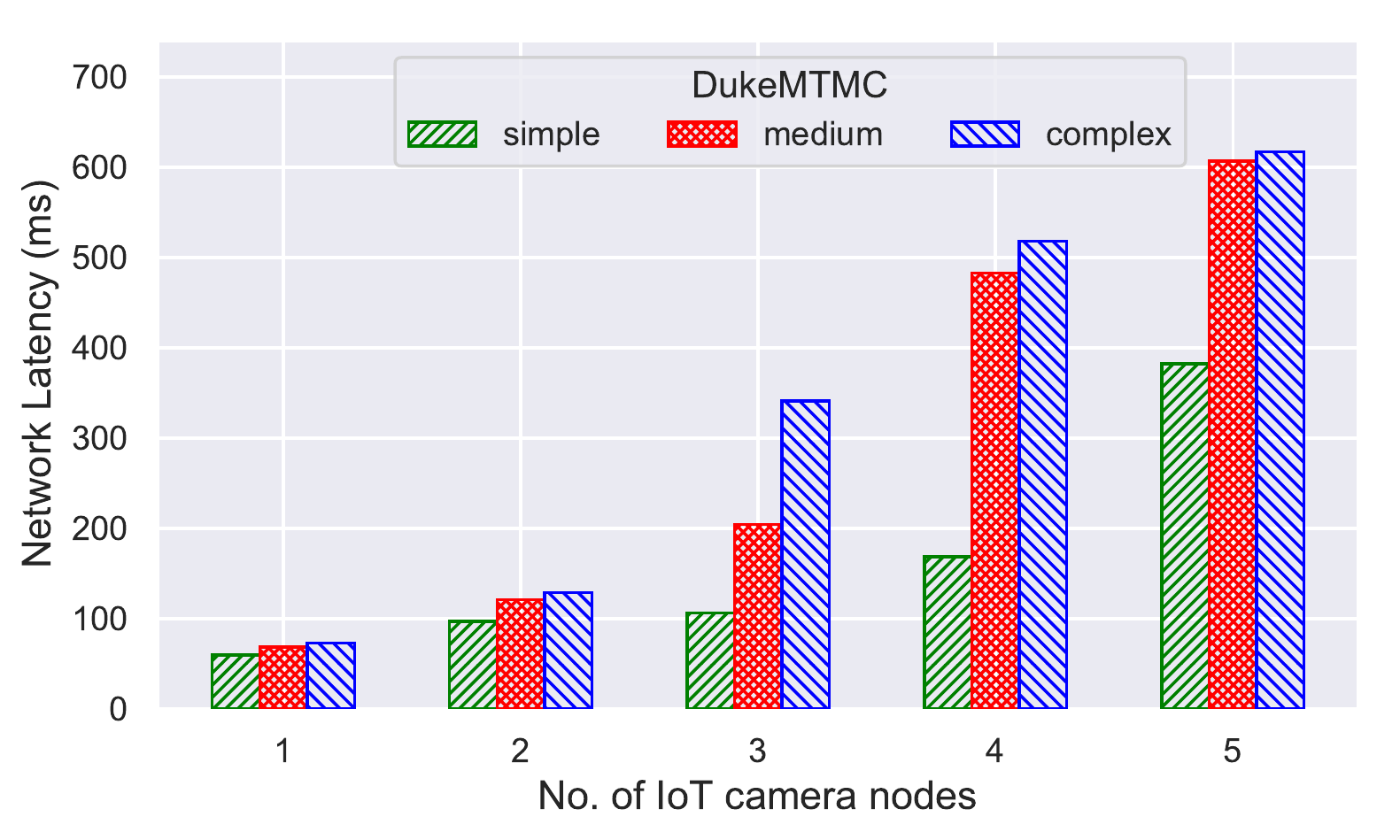}\label{fig:lat_vs_num_nodes_scene_dyn_duke}}
\caption[Latency vs. number of Edge nodes for scene dynamics.]{
Characterization of the impact of peer interference on the video frame transfer latency for frames with different scene dynamics from (a) JAAD and (b) DukeMTMC datasets. For complex scene dynamics a 5.6x and 8.4x increase in latency is observed for the JAAD and DukeMTMC dataset respectively.}
\label{fig:lat_vs_num_nodes_scene_dyn}
\end{figure}

\begin{table}
    \footnotesize
    \centering
    \caption{Summary of the impact of video frame size on network latency for JAAD and DukeMTMC workloads with simple (S), medium (M) and complex (C) scene dynamics (SD) video frames. All latency measurements are at the 95th percentile with video frames transmitted at 5 fps. $ONE_{Lat}$ is the per frame network latency to the Edge server when only the test node is active. $FIVE_{Lat}$ is the per frame network latency when the node under test and the 4 peer Edge nodes transmit video frames to the Edge server.} 
    \label{table:latency_summary}
    \renewcommand{\arraystretch}{1.2}
    \begin{tabular}{|p{2.2cm}|p{0.8cm}|p{0.8cm}|p{0.8cm}|p{0.8cm}|p{0.8cm}|p{0.8cm}|}
        \hline
        Dataset & \multicolumn{3}{c|}{JAAD} & \multicolumn{3}{c|}{DukeMTMC} \\
        \hline
		SD   & S   & M    & C   & S   & M  & C\\
        \hline
        $Size_{med}$ (KB)   &  610 & 760   & 970  & 1390 & 1670 & 1740\\
		\hline
		$ONE_{Lat}$ (ms)   &  32.09 & 35.16   & 46.09  & 59.71 & 68.73 & 72.72\\
		\hline
		$FIVE_{Lat}$ (ms)   &  150.28  &  164.56   & 262.43  & 382.47 & 606.98 & 617.16\\
		\hline
		$FIVE_{Lat}/ONE_{Lat}$ &  4.6x  &  4.6x   & 5.6x  & 6.4x & 8.8x & 8.4x\\                
		\hline
    \end{tabular}
\end{table}

Figure \ref{fig:lat_vs_num_nodes_scene_dyn} shows the per frame network latency measured at the test Edge node as the number of IoT camera nodes transmitting video frames (with simple, medium and complex scene dynamics) is increased from 1 to 5. Table \ref{table:latency_summary} summarizes the latency measurements. $ONE_{Lat}$ is the per frame network latency to the Edge server when only the test node is active. $FIVE_{Lat}$ is the per frame network latency when the node under test and the 4 peer Edge nodes transmit video frames to the Edge server.  We note that for video frames with complex scene dynamics, the ratio $FIVE_{Lat}/ONE_{Lat}$ is 5.6x for the JAAD dataset, and 8.4x for the DukeMTMC dataset.

\begin{table}
\footnotesize
\centering
\caption{Summary of network latency vs. frame rates (5 and 15 fps) and distance from Edge server (6m and 12m) for DukeMTMC complex scene dynamic video frames.}
\label{table:frame_rate_summary}
\renewcommand{\arraystretch}{1.2}
\begin{tabular}{|p{0.6cm} | p{0.8cm} | p{0.8cm} | p{0.9cm}|}
\hline
\multirow{3}{0.4cm}{Num. nodes} &  \multicolumn{3}{c|}{Network latency (ms)} \\\cline{2-4}
                            
                            & 5fps (at 6m) & 15fps (at 6m) & 5fps (at 12m)\\
\hline
1 & 72.72 & 80.60 & 96.35\\
\hline
2 & 128.97 & 409.82 & 162.15\\
\hline
3 & 341.18 & 438.01 & 390.75\\
\hline
4 & 518.31 & 585.58 & 526.95\\
\hline
5 & 617.16 & 631.76 & 657.88\\
\hline
\end{tabular}
\end{table}

We also investigate the impact of peer node interference at higher video frame rates and with increasing distance of IoT camera nodes from the Edge server. Table \ref{table:frame_rate_summary} compares the network latencies between 5 and 15 fps for complex scene dynamics video frames from the DukeMTMC dataset at both 6m and 12m. We note that the $FIVE_{Lat}$ at 15 fps is 1.02x higher for DukeMTMC compared to 5 fps, and at 12m is 1.06x higher compared to 6m.

The measurement results indicate that in an IoT machine vision application with multiple cameras transmitting video frames to the Edge server, a significant rise in network latency is observed at each IoT node as the number of peer nodes scale. Additionally, factors affecting latency include scene dynamics; frame rate, and distance of IoT camera nodes from the Edge server are less significant. In a real-world deployment, additional external interference effects from unrelated transmitters in the neighborhood of the deployment worsen the latency. Moreover, the network latency is dynamic due to scene changes (simple to complex), and the intermittent nature of external interference. 

\subsection{Evaluating the impact of video frame quality on network latency}
\label{sec:quality}

We investigate the impact on network latency when video frames with degraded quality are transferred from the IoT camera node to the Edge server. The degradation is caused due to discarding of information from the the video frame, resulting in a lower video frame size that can be potentially transmitted at reduced network latency. However, the lower frame size could adversely impact the accuracy of the machine vision application as well. The impact on the accuracy is application specific and will be evaluated in the context of a specific application in Section \ref{sec:accuracy}.

\subsubsection{\textbf{Video frame quality tuning knobs}}
\label{sec:tuningknobs}
We use the open source computer vision library OpenCV \cite{OpenCV} to explore different lossy image transformation techniques that can be applied to video frames to modify the frame size. We choose 5 such transformation techniques (which we call tuning knobs \cite{latency_ctrl}). These are described below:

\begin{enumerate}

\item \textbf{Knob1 - Resolution:} Video frame size can be reduced by decreasing its resolution while keeping the aspect ratio constant. The cv2.resize() function from OpenCV downscales an image to the specified resolution. We choose the resolutions 1312x736, 960x528, 640x352, and 480x256 as possible knob settings. Modifying resolution can reduce the video frame size by as much as 84\%. 

\item \textbf{Knob2 - Colorspace modifications:} Video frames can be converted from one colorspace to another (using cv2.cvtColor() function from OpenCV) resulting in total size reduction. There are more than 150 color-space conversion methods available in OpenCV. We choose BGR$\leftrightarrow$Gray, BGR$\leftrightarrow$HSV, BGR$\leftrightarrow$LAB and BGR$\leftrightarrow$LUV colorspace modifications as possible knob settings. Our choice of color space modifications can reduce the video frame size by as much as 62\%. 

\item \textbf{Knob3 - Blurring:} Video frames can be blurred by passing them through various low pass filters. The cv2.blur() method from OpenCV blurs an image using normalized box filter. We choose filter kernel sizes of (5,5), (8,8), (10,10) and (15,15) as possible knob settings. Blurring the video frames can reduce the video frame size by as much as 46\%.

\item \textbf{Knob4 - Artifact removal:} For cameras mounted at fixed positions, the background of recorded video stream is largely static over consecutive frames in the video. Thus video frame size can be reduced by removing the static background with stationary artifacts in it. First setting of this knob uses motion detection to detect and preserve moving objects in video frames, as well as to perform background subtraction to remove all the stationary objects from the video frames. In the second setting of this knob, we detect moving objects in the video frames and retain only their contours. Knob4 uses a combination of the OpenCV functions cv2.absdiff(), cv2.threshold(), cv2.dilate(), cv2.findContours() to perform the above video frame modifications. Removing artifact information can reduce the video frame size by as much as 98\%.


\item \textbf{Knob5 - Frame differencing:} We applied frame differencing (using cv2.absdiff() function from OpenCV) on pixel values between pairs of consecutive video frames to selectively drop frames. We hypothesize that dropping video frames with similar content (within a threshold) will not adversely affect the machine vision task. We choose 5 knob values ranging from 0 to 0.72, where 0 represents pixel wise identical frames, and 1 represents completely dissimilar frames. For a stream of 100 simple dynamics images from the JAAD data set, this knob reduces the median image size by up to 40\%.

\end{enumerate}

\begin{figure}
    \center{\includegraphics[width=0.7\textwidth]{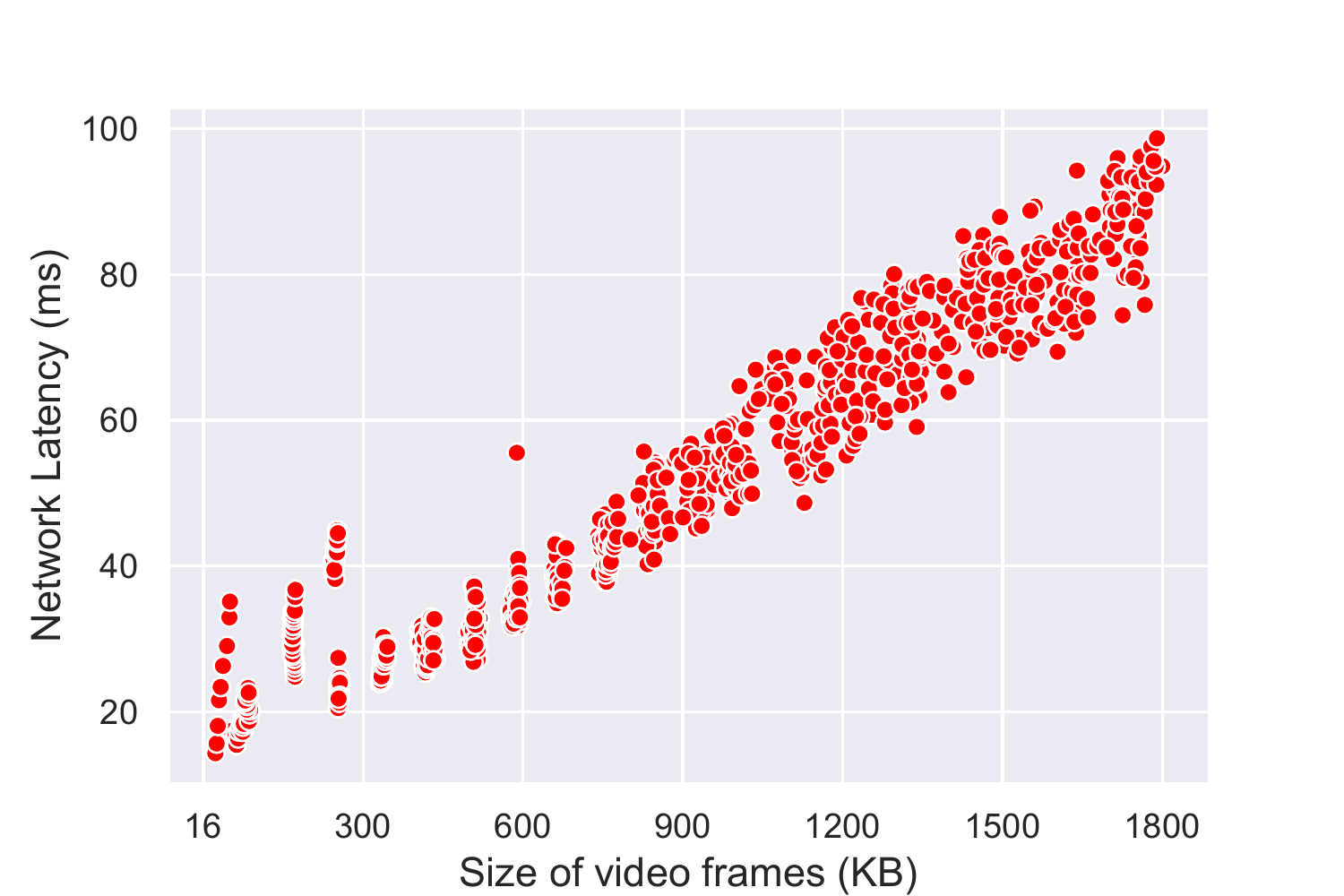}}
    \caption{\label{fig:latVSsize} Network latency vs. video frame size. Video frame sizes are obtained by the application of different combinations of the 5 tuning knobs that modify the image quality.}
\end{figure}

The application of combinations of the 5 tuning knobs identified above result in different sizes of video frames, all lower than the original. Figure \ref{fig:latVSsize} shows the resulting impact on the network latency (95th percentile) from the IoT camera node to the Edge server. The measurements were done by applying multiple tuning knob combinations (935 in all) to video frames drawn from the JAAD and DukeMTMC dataset. From Figure \ref{fig:latVSsize} we note that the Wi-Fi transmission latency shows an approximately linear variation with video frame size. A 4x reduction in video frame size could potentially yield a 4x reduction in wireless network latency.  We also note that multiple knob combinations map to the same video frame size (and hence network latency). However, these knob combinations could result in different application accuracy - which we characterize in the next section.

\subsection{Evaluating impact of video frame quality on the accuracy of pedestrian detection machine vision application}
\label{sec:accuracy}

While we note the ability of image tuning knobs to reduce network latency by reducing the size of the video frames, the question remains as to the impact of the lower sized video frames on the accuracy of the machine vision task. In general, the impact is dependent on the particular machine vision application. We evaluate the impact on pedestrian detection application accuracy using OpenPose with video frames drawn from the JAAD and DukeMTMC datasets. The OpenPose project from CMU \cite{openpose1, openpose2, openpose3, openpose4} is an open source real-time multi person system to detect human body, hand and facial keypoints ((x,y) coordinates of different body parts) on individual images. We input the original and modified video frames from JAAD and DukeMTMC dataset with simple, medium and complex scene dynamics to OpenPose to generate the pose detected video frames and keypoint locations. From these keypoints, bounding boxes are created for each detection with the top-left and bottom-right most coordinates. A set of resulting bounding boxes is presented as the final output. In order to evaluate these detections, each ground truth bounding box for that frame (available for the two datasets) is matched exclusively to the outputted bounding box based on highest Intersection over Union (IoU) overlap. Positive matches with an IoU greater than a threshold are considered True Positives; result bounding boxes without ground truth matches are considered False Positives; and each unmatched ground truth box is considered a False Negative. These records are utilized for the F1 score calculation.

For pedestrian detection, we utilize the F1-score metric with an Intersection-over-Union (IoU) threshold of 0.5 as the application accuracy metric. Equation \ref{F1} defines the calculation for F1. Precision is $\frac{TP}{(TP+FP)}$ and Recall is $\frac{TP}{(TP+FN)}$, where $TP$, $FP$, and $FN$ are the number of True Positives, False Positives, and False Negatives respectively. 

\begin{equation}\label{F1}
    \text{F1} = 2 \times \frac{Precision \times Recall}{Precision + Recall}
\end{equation}{}

\begin{figure}
\centering
\subfloat[]{\includegraphics[width=0.49\textwidth, keepaspectratio
]{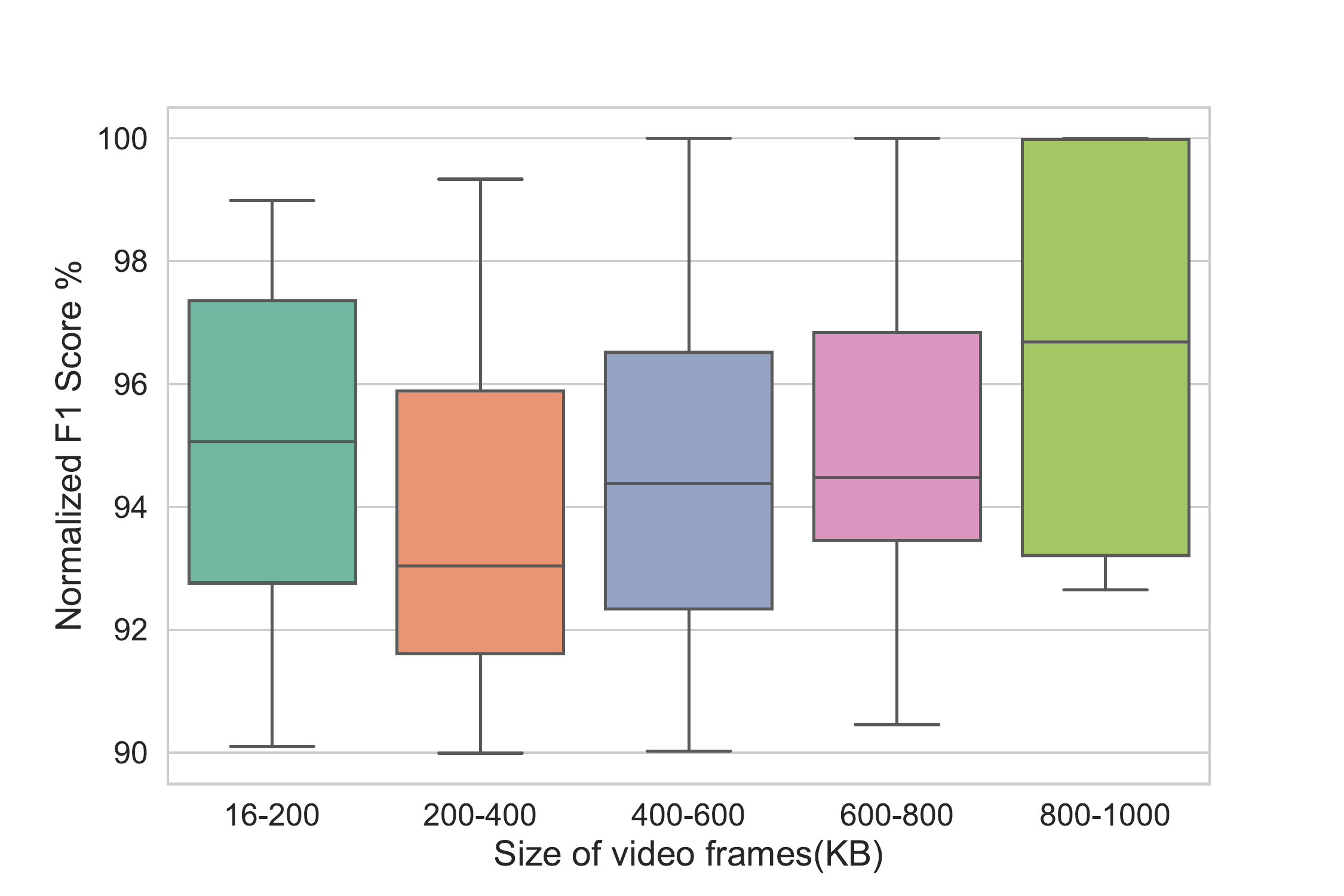}\label{fig:jaad_box}}\hspace{2pt}
\subfloat[]{\includegraphics[width=0.49\textwidth, keepaspectratio]{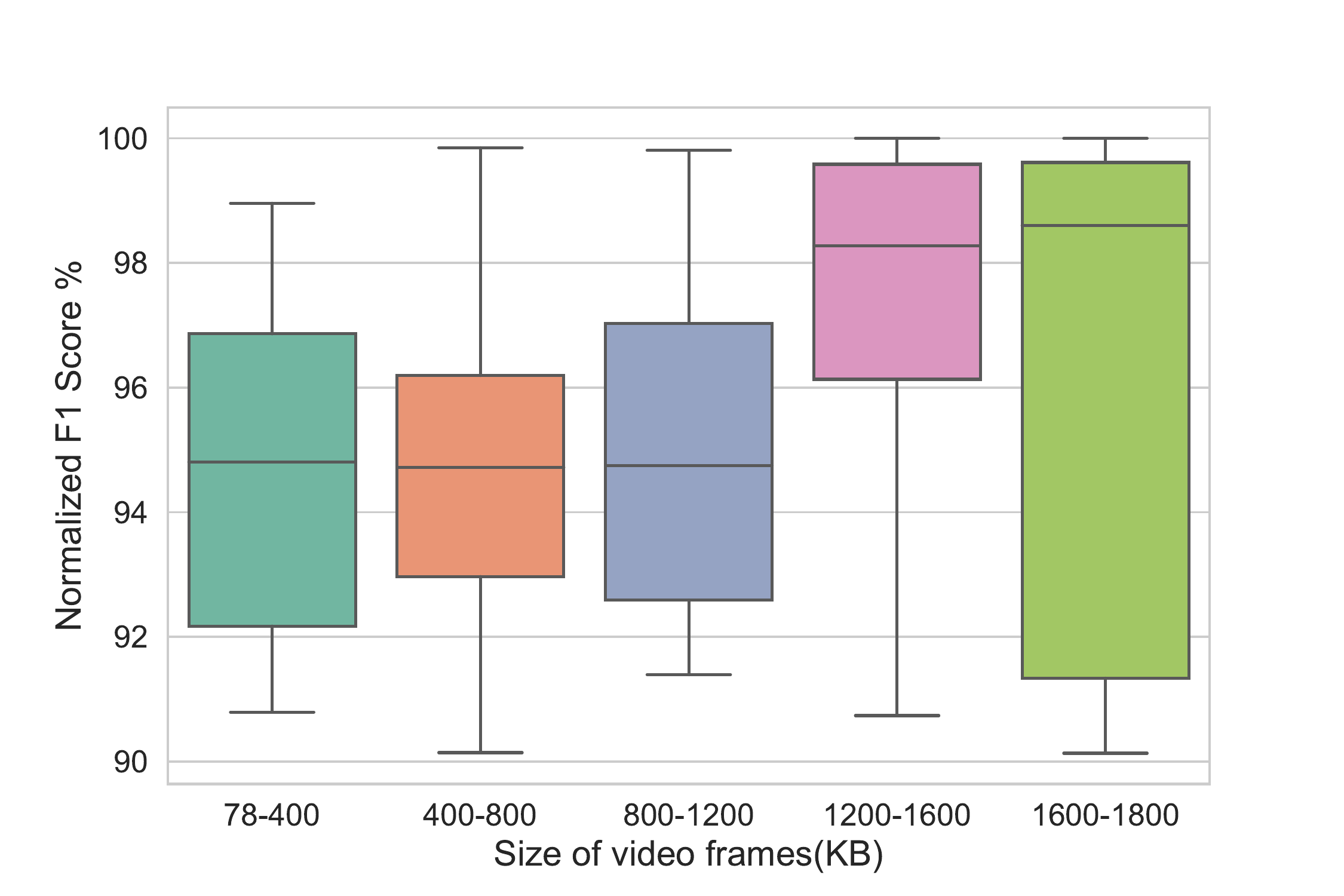}\label{fig:duke_box}}
\caption[Accuracy vs. image size.]{
Normalized F1 expressed as a percentage for Openpose pedestrian detection application from (a) JAAD and (b) DukeMTMC datasets. Note that each video frame bucket corresponds to different combinations of the knob settings with different resulting accuracy.}
\label{fig:accu_vs_size}
\end{figure}

We evaluate the impact of the tuning knobs on the video frame size and pedestrian detection accuracy (F1) for JAAD and DukeMTMC datasets. To do this, we first calculate the F1 score for modified video frames (for all knob combinations) and normalize it with the baseline F1 score of unmodified video frames. Figure \ref{fig:accu_vs_size} shows the plot of the normalized F1 expressed as a percentage vs. video frame size for JAAD and DukeMTMC datasets. The video frame size buckets in Figure \ref{fig:accu_vs_size} corresponds to different combinations of the knob settings with different resulting accuracy. Note that higher F1 indicates higher accuracy. We have excluded knob combinations with resulting accuracy of less than 90\%. This reduces the total knob combinations to 159 for JAAD and 140 for DukeMTMC. Further, due to the computationally intensive nature of knob 4, we exclude knob 4 to maintain the image modification overheads to under 10 ms, bringing the total knob combinations to 70 for JAAD and 92 for DukeMTMC.

From the measurements in Sections \ref{sec:quality} and Sections \ref{sec:accuracy} we note the possibility of compensating for the increased network latency in the presence of channel interference through reducing the video frame quality just sufficiently, such that application accuracy demands are met (where feasible). This is an example of the paradigm of approximate computing - where despite computational approximations (video frame quality in our case) acceptable performance (application accuracy) can be obtained while gaining on another performance metric (network latency). Note that the limits of the trade-off needs to be determined and characterized by the application developer for a particular application of interest. We utilize the above observed trade-off between network latency and application accuracy in designing a network latency controller. Under dynamically varying network latency conditions, Mez uses this latency controller to automatically adjust video frame quality such that the application specified network latency, and accuracy bounds are met.


\section{Mez API and Architecture}
\label{sec:mez_architecture}

In this section we present the Mez API that publishers and subscribers use to interact with Mez. We then provide an overview of the data model supported by Mez, and describe the architecture of Mez. 

\subsection{Mez API}
\label{sec:mez_apis}

\begin{figure}[h]
  \begin{center}
    \includegraphics[width=0.8\textwidth]{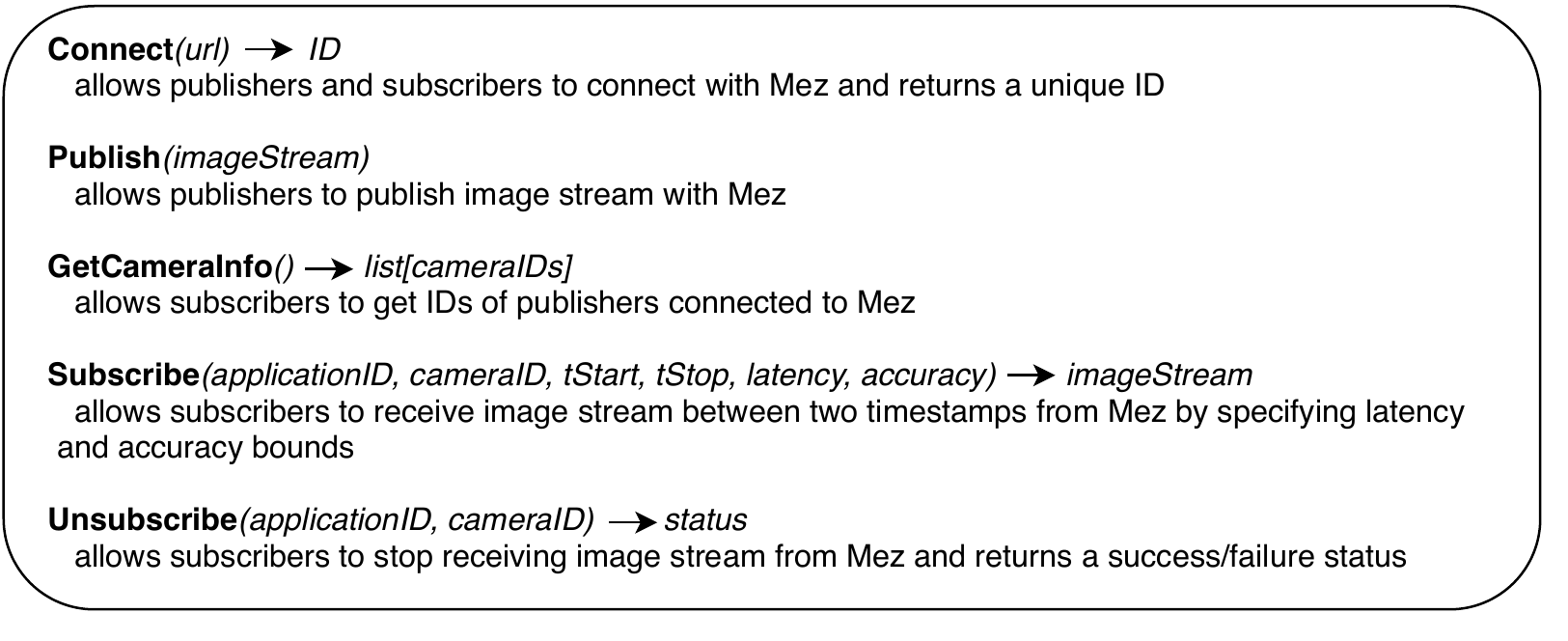}
  \end{center}
  \caption{A summary of the API provided by Mez}
  \label{fig:Mez_apis}
\end{figure}

Mez has a simple API interface consisting of 5 API calls. As shown in Figure \ref{fig:Mez_apis} - the APIs are Connect, Publish, GetCameraInfo, Subscribe, and Unsubscribe.

Connect API is used by publishers (IoT camera nodes) and subscribers (machine vision applications) to connect to Mez. Publishers and subscribers are assigned a unique ID by Mez. Publish API allows publishers to push a stream of time stamped video frames to Mez. The GetCameraInfo API is used by subscribers to discover publishers. Subscribe API is used by subscribers to receive streaming video frames generated by a specific publisher. Additionally, the subscribers can specify begin and end times for the subscription, along with the desired latency and accuracy bounds for the video stream. Note that the end time could be in the future, in which case Mez delivers video frames as they become available. The Unsubscribe API is used by subscribers to terminate an ongoing subscription.

\subsection{Data Model}
\label{sec:data_model}

Mez’s data model is a simple key-value pair. The keys are the timestamps of a video frame, and the values are individual video frames. Video frames are stored in Mez in the same chronological order in which they are received from the publisher. Mez supports at-most-once delivery of video frames to the subscriber to limit the bandwidth consumption on the wireless channel. Any resend requests need to be done by the subscriber at the application level, since only the application can determine if a resend is needed considering task deadlines, and redundancy in the video frames. 

\subsection{Mez system architecture}
\label{sec:log_arch}

The Mez system model consists of multiple IoT camera nodes, and an Edge server connected by a wireless network.  The machine vision applications run on the Edge server, which has considerably more processing power than the IoT camera node.

\begin{figure}[h]
  \begin{center}
    \includegraphics[width=0.8\textwidth]{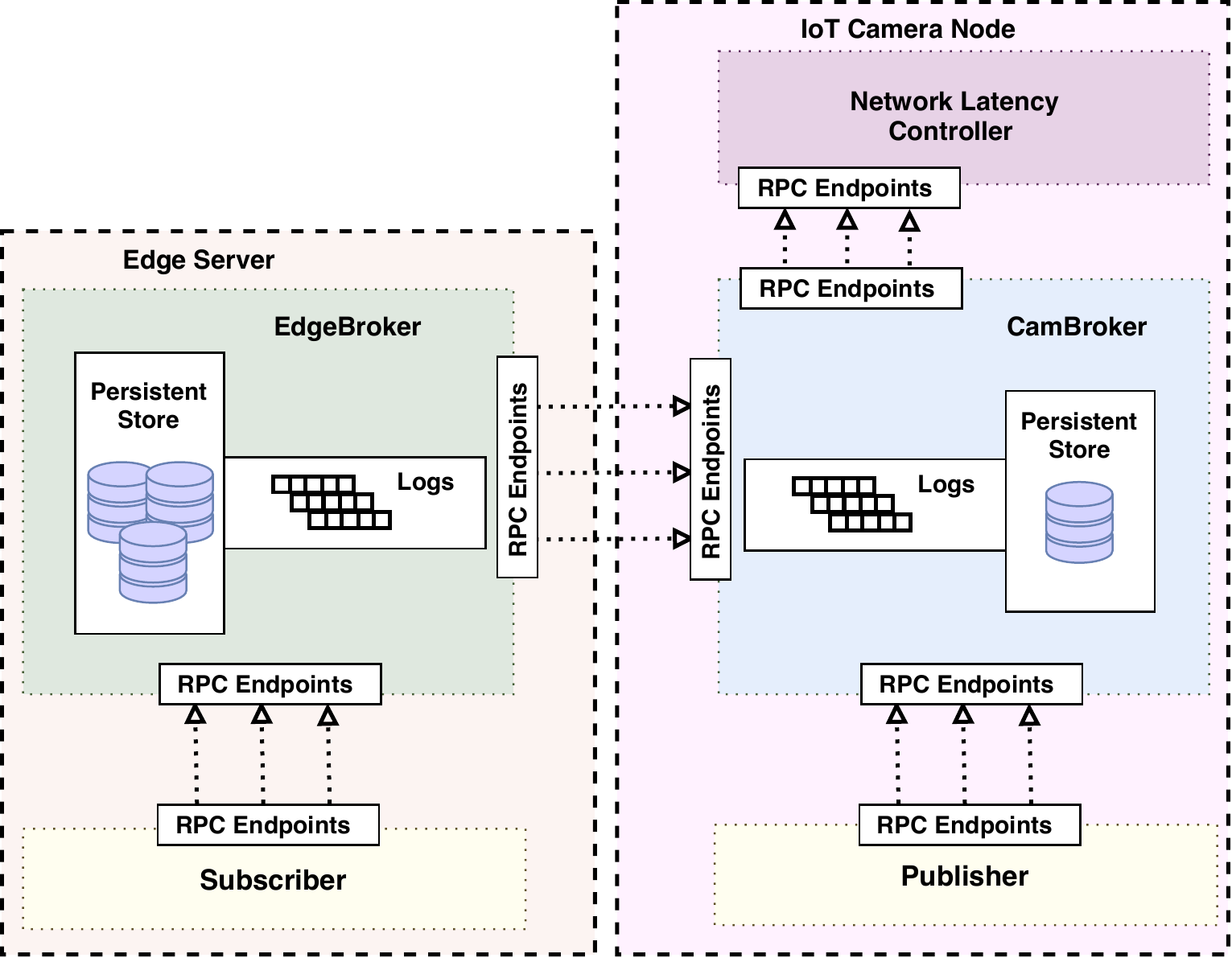}
  \end{center}
  \caption{Detailed architecture of Mez}
  \label{fig:arch}
\end{figure}

As shown in Figure \ref{fig:arch}, Mez consists of 3 components - a message broker, an in-memory log, and a network latency controller. The message broker implements the Mez API, the in-memory log is used to store video frames, and the network latency controller monitors  wireless channel conditions, automatically adjusting the video frame quality to meet the application specified latency, and accuracy requirements.

A key architectural feature of Mez is the replication of the in-memory logs between the IoT camera nodes and the Edge server. The timestamped video frames generated by the publisher are initially stored in the in-memory log associated with the camera node. Maintaining the log at the IoT camera node allows buffering of video frames during conditions of intermittent connectivity with the Edge server.  Upon a subscription request to the video stream by the application, video frames are transferred from the IoT camera node log to the Edge server log. The network controller resides on the IoT camera nodes, and modifies the video frames transmitted to the Edge server to satisfy the latency-accuracy requirements of the subscriber. The on-demand transfer of video frames reduces channel interference by limiting unneeded transmission in the wireless channel. Additionally, the reduced transmission also serves as a power saving opportunity at the IoT camera node. The in-memory logs are persisted on durable storage (SSD/disk) on both the IoT camera nodes, and the Edge server. However, to minimize storage latency, all requests are served from the in-memory log. The persistent storage is only used to reconstruct the in-memory log during reboot after node failure (see Section \ref{sec:mez_faults}).


\section{Mez design}
\label{sec:mez_design}

In this section, we present the detailed design of the different components of Mez - brokers, in-memory log, and the network latency controller. 

\subsection{Brokers}
\label{sec:brokers}
The brokers implement the Mez API, and interface with the log storage. Independent brokers are present on the Edge server and the IoT camera node. The broker on the Edge server (EdgeBroker) implements all the APIs shown in Figure \ref{fig:Mez_apis} except Publish. Additionally, it implements two internal APIs, Register, and Unregister for IoT camera nodes to register/unregister with the Edge Server. The broker on the IoT camera node (CamBroker) implements all the APIs shown in Figure \ref{fig:Mez_apis} except GetCameraInfo. The CamBroker also interfaces with the network latency controller on the IoT camera node. All APIs are implemented using gRPC \cite{grpc} with TLS for authentication and encryption of data in transit.

\subsection{Network Latency Controller}
\label{sec:controller}

\begin{figure}[h]
    \center{\includegraphics[width=1\textwidth]{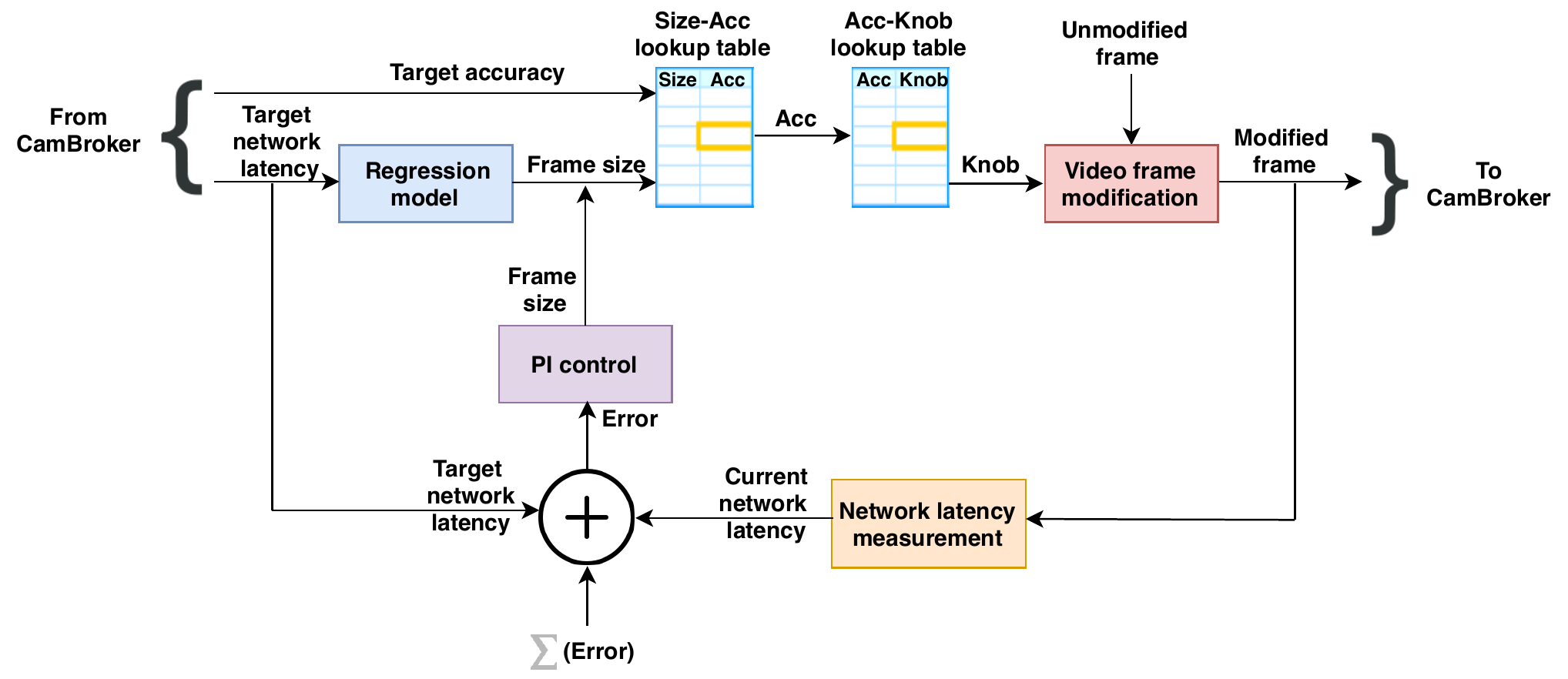}}
    \caption{\label{fig:latency_controller} Block diagram of the network latency controller}
\end{figure}

Figure \ref{fig:latency_controller} shows the block diagram of the latency controller. The CamBroker sends the latency and accuracy demands (target network latency and target accuracy) from the consumer application to the latency controller through an internal SetTarget API call. 

The video frames published at the CamBroker are sent to the latency controller through a Control API call. Note that the latency controllers on each IoT camera node operate independently of one another. The lack of centralized control allows the scaling of the IoT camera nodes. 

The control algorithm is outlined in the pseudo code shown in Algorithm \ref{controlalgo}. The network latency for different video frame sizes (see Section \ref{sec:quality}) are assumed to be available from prior characterization of the video frames in the targeted deployment environment. The application accuracy for different tuning knob settings is also assumed available for the targeted application through prior characterization. The almost linear dependence of latency on image size observed in Section \ref{sec:tuningknobs} facilitates the use of linear regression model of latency on video frame size. 

\begin{algorithm}[h]
\SetAlgoLined
\KwResult{Image quality knob setting }
 latencyTarget\;
 accuracyTarget\;
 errorThreshold\;
 nominalImageSize $\leftarrow$ RegressionModel(latencyTarget)\;
 latencyError $\leftarrow$ latencySampled - latencyTarget\;
 \While{latencyError $>$ errorThreshold}{
  imageSize = nominalImageSize + K1*latencyError +  K2*latencyErrorIntegral\;
  accuracy $\leftarrow$ BinarySearchTree.search(imageSize)\;
  knobSetting $\leftarrow$ HashTable.lookup(accuracy)\;
  \eIf{accuracy $>$ AccuracyTarget}{
   return knobSetting\;
   }{
   return(No feasible solution)\;\
  }
  latencyError $\leftarrow$ latencySampled - latencyTarget\;
 }
\caption{Latency control algorithm}
 \label{controlalgo}
\end{algorithm}

The control is implemented in two steps - In Step 1, the error (error and integral of error for Proportional-Integral control) between the observed and the specified latency is used to determine the largest video frame size that can potentially satisfy latency requirements. K1 (proportional) and K2 (integral) in Algorithm \ref{controlalgo} are the PI controller tuning parameters. In Step 2, the frame size is then used as a key in a look-up table to determine the associated application accuracy. A secondary look-up table uses the application accuracy obtained from the first lookup as key to determine the corresponding tuning knob settings. The lookup tables are implemented with Binary Search Trees and Hash tables to facilitate efficient queries. The video frames transmitted from the IoT camera node to the Edge server are modified based on the knob combinations using OpenCV libraries. The network latency is measured again at the next sampling interval, and if the error exceeds a preset threshold, Steps 1 and 2 are repeated. 

If the application requested latency and accuracy are infeasible, the application is notified. At this point, the application has to decide whether to continue operation with relaxed latency/accuracy requirements, or notify the system operator of failure.

\subsection{In-memory Log}
\label{sec:mez_storage}

\begin{figure}[h]
  \begin{center}
    \includegraphics[width=0.7\textwidth]{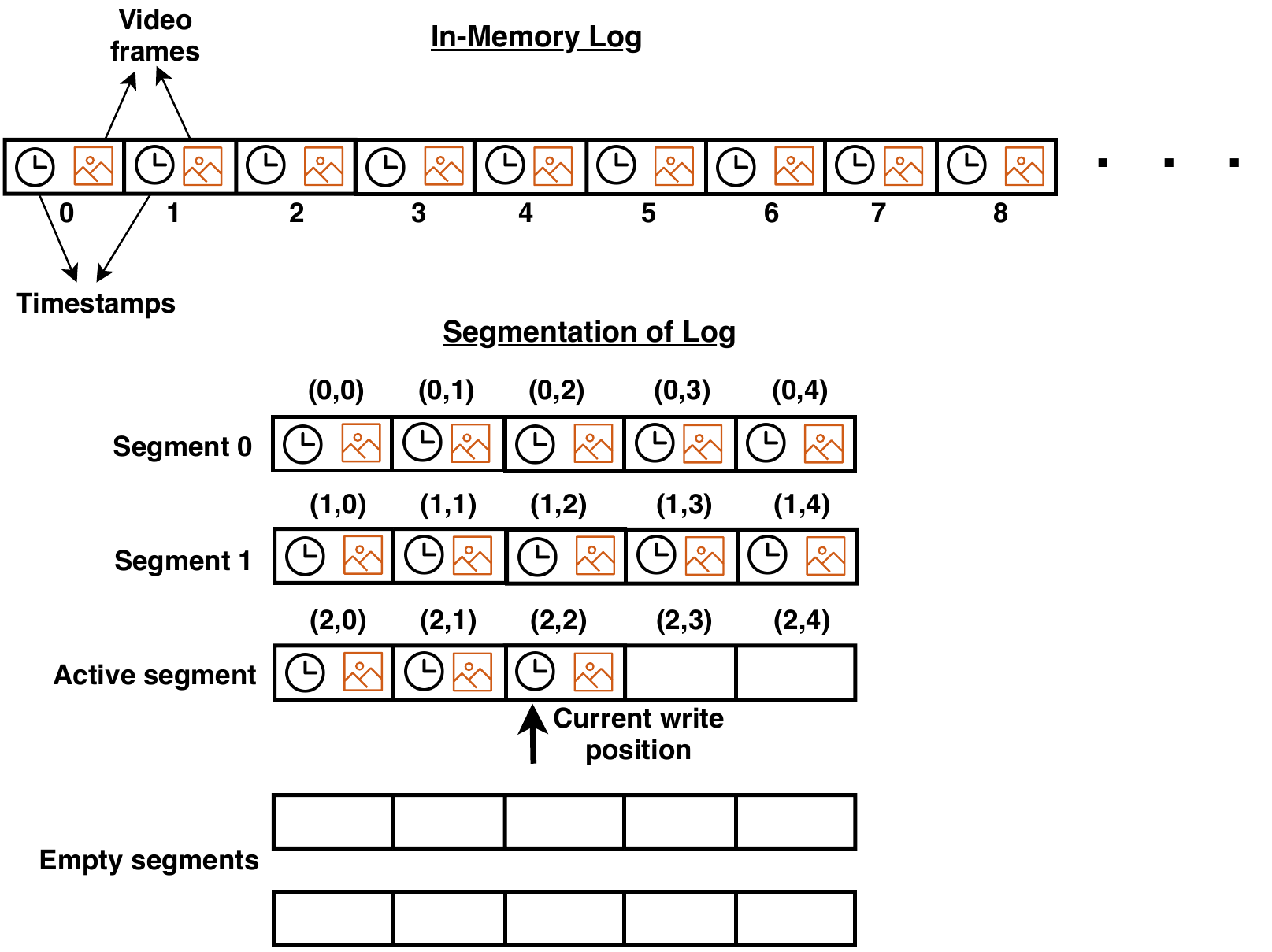}
  \end{center}
  \caption{Mez in-memory log. The storage is an in-memory log, which is an append-only, circular buffer.  The log consists of <timestamp, video frame> key-value pairs stored in increasing order of timestamps. Concurrent read/write performance is improved through fine grained locking by segmenting the log.}
  \label{fig:storage_arch}
\end{figure}

The design of Mez storage is targeted to provide low latency read-write operations. We take advantage of the particular features of machine vision at the IoT edge, both to ensure high performance, and simplify the design of the storage. As shown in Figure \ref{fig:storage_arch}, the storage is an in-memory log, which is an append-only, circular buffer.  The log consists of <timestamp, video frame> key-value pairs stored in increasing order of timestamps. The log at the IoT camera node stores video frames either generated from a single camera, or those modified by the latency controller to satisfy latency requirements. This log is in turn replicated on demand at the Edge server. For N IoT camera nodes, the Edge server thus holds N replicated logs. Subscriber machine vision applications are served from the log at the Edge server. A machine vision application subscribes to one or more such logs. Also, one or more machine vision applications could subscribe to a particular log. The logs are hence designed to support a single writer, but multiple readers. 

To ensure low latency, the logs only utilize the DRAM for storage. Unlike general purpose storage, there is no requirement to delete arbitrary video frames from the log. Instead, video frames with increasing time stamps are appended to the log, which wraps back when the capacity is exceeded, overwriting existing entries with older timestamps. An attempt to append a video frame with a timestamp earlier than the last entry in the log is rejected. The lack of a need to support update and delete operations, and the sorted (by timestamps) video frames in the log, simplify the design of the log and prevent memory fragmentation. Point queries are done efficiently with binary search. Range queries are also readily supported by querying the starting and ending timestamp, returning the video frames corresponding to an interval that includes the requested time range. 

 A 1 GB of in-memory log at the IoT camera node holds approximately 7 minutes worth of video frames (assuming 500 kB per frame, and 5 fps). The real-time machine vision applications are assumed to consume the data within this time frame. The log is persisted on the disk (in the background) only for recovery from failure (described in Section \ref{sec:mez_faults}). Due to the possible physical insecurity of the hardware, the video frames stored on disk are encrypted at rest. The encryption/decryption, and disk accesses are relatively long latency operations, motivating the avoidance of disk access in the read/write critical path. For the short time duration the video frames are held in the DRAM, the data is assumed to be safe from illegal access due to the volatile nature of the memory.  

Although the log is replicated from the IoT camera node to the Edge server, no attempt is made at the storage layer to ensure consistency of the video frames between the IoT node and the Edge server. Instead, we rely on the lower layers of the network (TCP) for accurate replication. Also, in practice we observe that the ability of machine vision applications to tolerate errors in the video frame data, allows the use of  simpler transport protocols (UDP) that does not support re-transmissions. Concurrent read/write performance is improved through fine grained locking by segmenting the log. Each segment is protected with read-write locks. Note that reads can occur from many segments concurrently, while only one segment is active for write.


\subsection{\textbf{Fault tolerance}}
\label{sec:mez_faults}

Mez is designed to recover from the following failures:

\begin{itemize}
    \item Crashes of brokers - EdgeBroker and CamBroker
    \item Crash of the network latency controller at the IoT camera node
    \item Corruption of log segments on disk 
\end{itemize}

\textbf{Fault detection: }
Mez uses RPC timeouts to detect failures. The EdgeBroker detects failed CamBrokers through  time outs on the Subscribe API call. The publisher detects failed CamBrokers through timeouts on the Publish API call. Subscriber applications detects a failed EdgeBroker through time outs on the Subscribe API call. The Cambroker detects a failed latency controller through time outs on the the internal SetTarget API call. The time out duration for the Subscribe RPC depends on the video frame rate (fps), baseline wireless network latency, and if TCP is used - the re-transmit timeout. The Publish timeout is typically small since the camera and the CamBrokers are connected through a high speed interface such as USB. The Control timeout is determined by the time taken by the controller to modify the video frame. We thus avoid use of explicit heartbeats, and instead take advantage of the continuous streaming of video frames from the cameras to detect component failures. For systems where the camera transmissions may be intermittent, explicit heartbeats will need to be added to monitor the health of individual components.

\textbf{Fault recovery: } When the subscriber application detects that the EdgeBroker has failed, it tries to reconnect with the EdgeBroker a finite (configurable) number of times or until it gets a response. As a part of the recovery process, the EdgeBroker reconstructs the logs persisted in the disk. A CRC is calculated and stored along with the on-disk log segments to detect partially written segments, which are discarded during the recovery process. The EdgeBroker then starts to accept connections from retrying subscriber applications. The CamBrokers follow a similar recovery process. Note that Mez has no inherent mechanism to restart failed brokers.  Instead, Mez relies on an external service such as Kubernetes \cite{kubernetes} to restart failed brokers (see Section \ref{sec:discussion}).


\section{Evaluation}
\label{sec:results}

We first evaluate the proposed network latency controller in isolation, followed by the evaluation of Mez integrating the controller. The IoT Edge test bed used in the evaluation is described in Section \ref{sec:testbed}. It consists of one Edge server, and five IoT camera nodes connected to the Edge server through Wi-Fi (802.11ac). All latencies are measured between one of the IoT camera nodes, and the Edge server. The workload used is the pedestrian detection application with OpenPose described in Section \ref{sec:accuracy} using video frames from the JAAD and DukeMTMC datasets described in Section \ref{sec:testbed}. All latencies are measured at the 95th percentile with video frames streamed from the IoT camera node to the Edge server at 5 fps. The latency is calculated as time difference $t_{Received} - t_{Send}$.  The Edge nodes and the Edge server are synchronized using PTP synchronization protocol \cite{ptp} before the start of the measurements.

\subsection{Latency Controller Evaluation}
\label{sec:controller_eval}

\begin{figure}[h]
  \centering
  \subfloat[]{\label{fig:lat_ctrl_jaad_complex}\includegraphics[width=62mm]{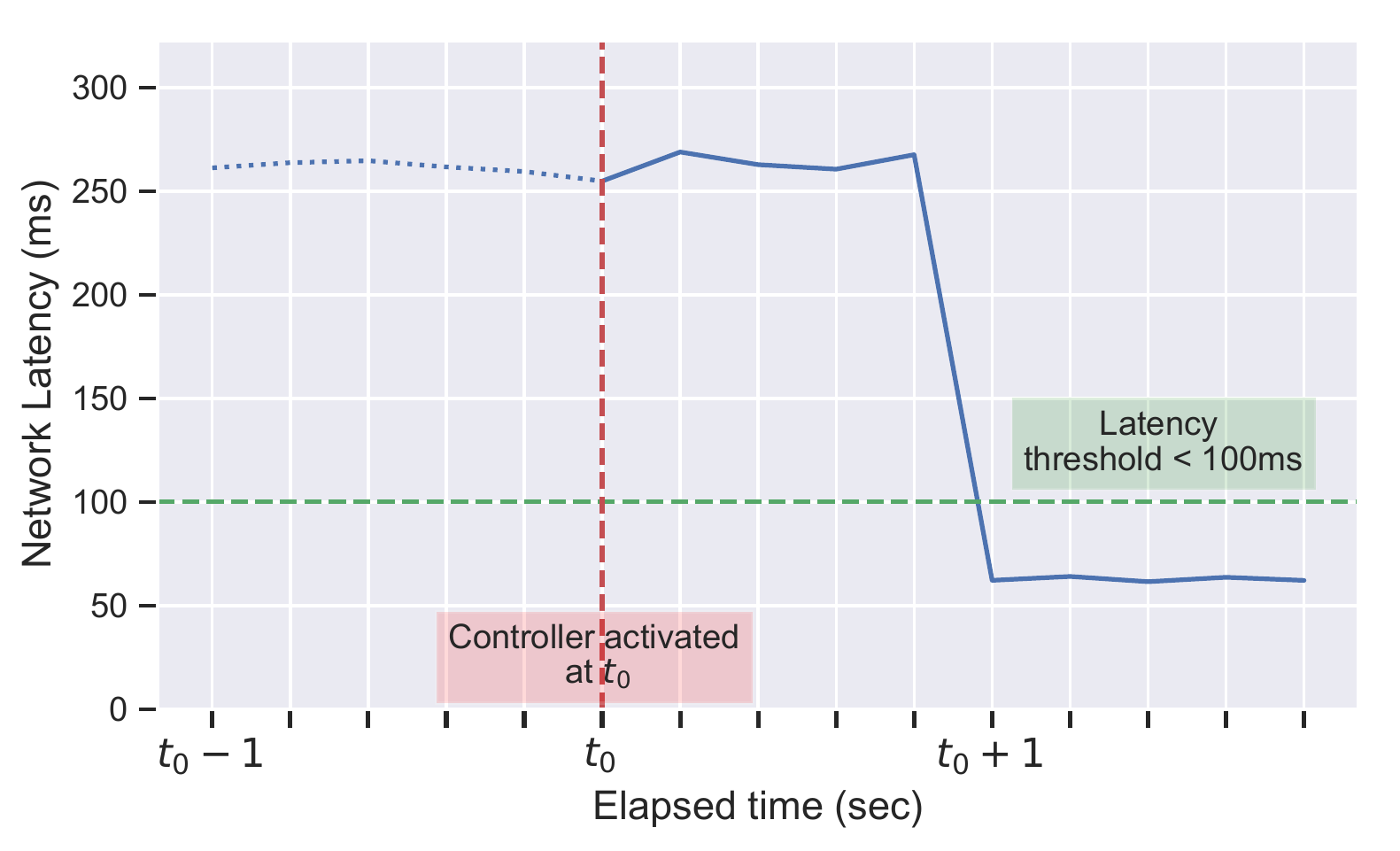}}
  \subfloat[]{\label{fig:lat_ctrl_duke_complex}\includegraphics[width=62mm]{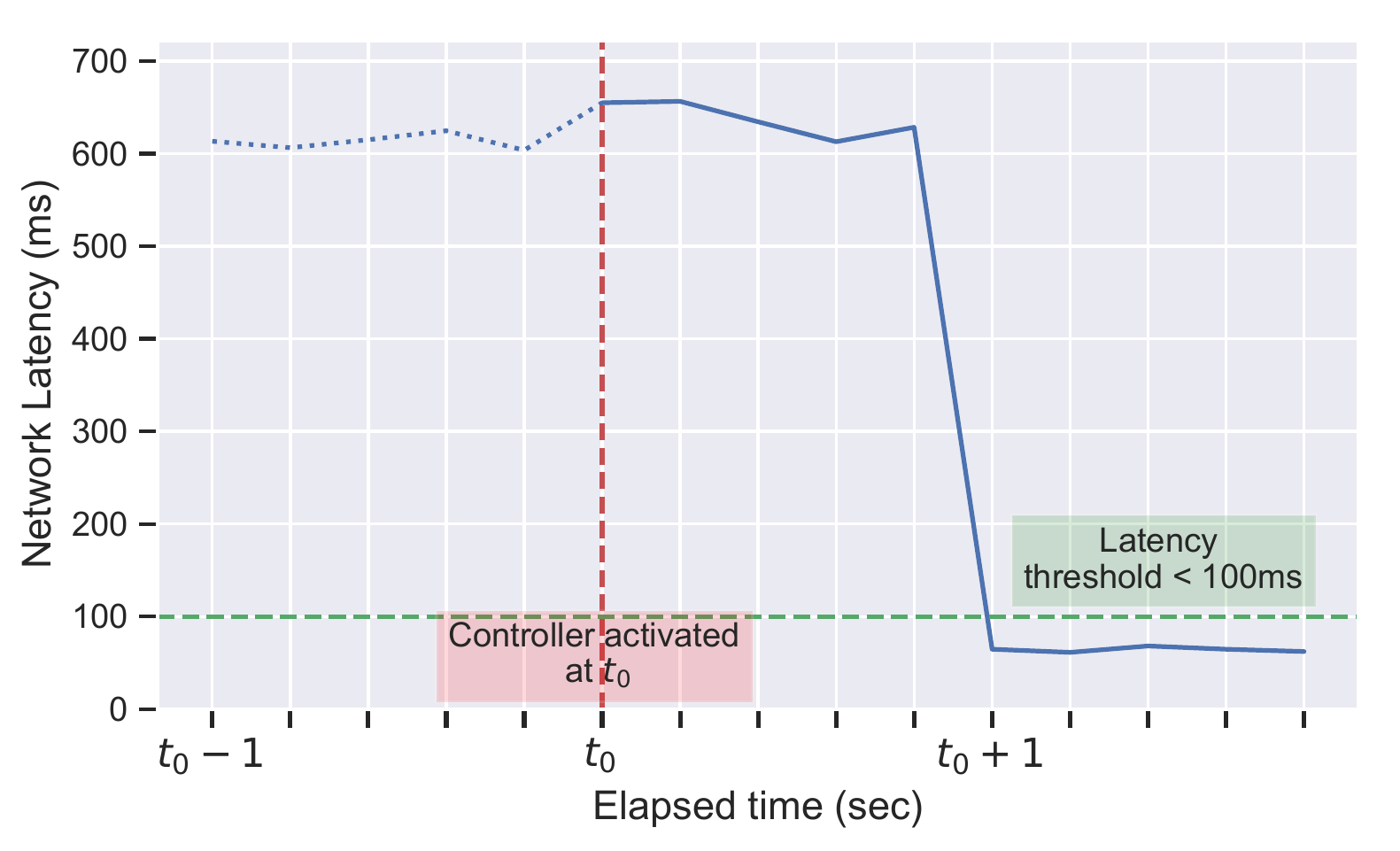}}
  \caption{Latency controller step response for JAAD (a) and DukeMTMC (b) complex scene dynamics video frames. The measurement is taken with one test IoT node, and 4 peer IoT nodes.}
  \label{fig:controller_plots}
\end{figure}

To evaluate the latency controller, the desired latency threshold is set under 100ms, and the desired application accuracy (normalized F1 score) is set above 95\%. Figure \ref{fig:lat_ctrl_jaad_complex}  shows the step response of the controller for complex scene dynamics video frames from the JAAD dataset. With no latency control, the median latency is 260 ms due to interference from the 4 peer Edge nodes. With the controller enabled, the median latency is less than the latency threshold of 100ms.  The controller is able achieve an application accuracy of above 96\%. Figure \ref{fig:lat_ctrl_duke_complex} shows the controller step response for the DukeMTMC dataset with complex scene dynamics. With no latency control, the median latency is 650 ms for complex scene dynamics due to interference from the 4 peer Edge nodes. With the controller enabled, the median latency is less than the latency threshold of 100ms. In all cases, the controller is able to achieve an application accuracy of above 95\% with a settling time of less than one second. Table \ref{table:controller_summary} summarizes the latency reduction and the resulting F1 score for JAAD and DukeMTMC datasets for all scene dynamics achieved using the proposed controller. 

\begin{table}[h]
\footnotesize
\centering
\caption[Summary of controller results for JAAD and DukeMTMC workloads]{Summary of  median video frame size after modification by the latency controller, ($Size_{med}$), Normalized F1 Score expressed as a percentage, 95th percentile latency reduction  with controller ($Lat_{red}$) for JAAD and DukeMTMC dataset for simple (S), medium (M) and complex (C) scene dynamics (SD).}
\label{table:controller_summary}
\renewcommand{\arraystretch}{1.2}
\begin{tabular}{|p{0.9cm} | p{0.62cm} | p{0.62cm} | p{0.62cm} | p{0.62cm} | p{0.62cm} |  p{0.62cm}|}
\hline
Dataset &  \multicolumn{3}{c|}{JAAD} & \multicolumn{3}{c|}{DukeMTMC}\\
\hline
SD & S   & M    & C   & S   & M  & C \\
\hline
$Size_{med}$ (KB) &  124 & 173  & 228  & 293 & 172 & 371\\
\hline
F1 score (\%) &  99.1  &  98.3 & 96.7  & 98.9  & 96.7 & 95.8\\
\hline
$Lat_{red}$ & 6.8x & 7.2x  & 4.1x  & 7.9x & 9.5x & 10.1x\\
\hline
\end{tabular}
\end{table}

\begin{figure}
  \centering
  \subfloat[]{\label{fig:F1_error_jaad}\includegraphics[width=58mm]{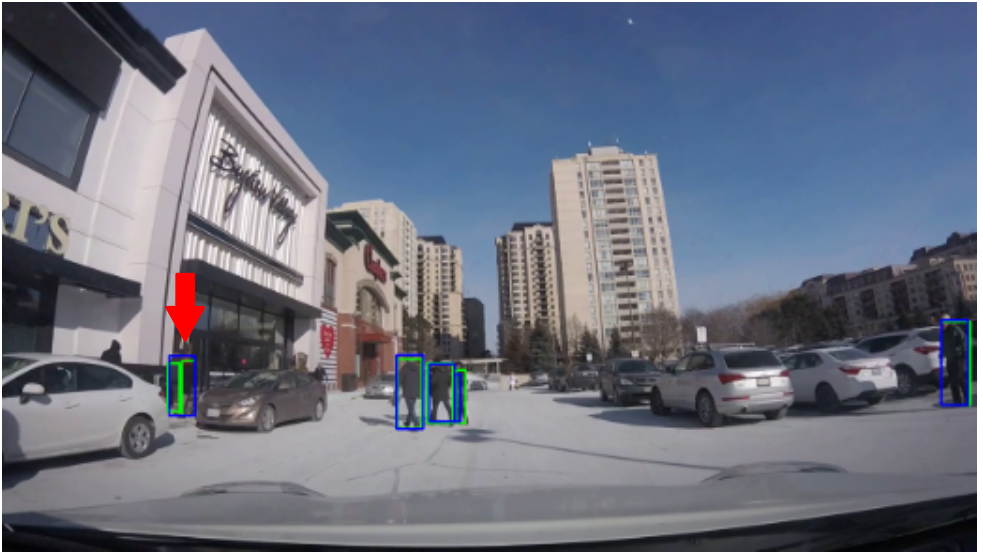}}
  \subfloat[]{\label{fig:F1_error_duke}\includegraphics[width=58mm]{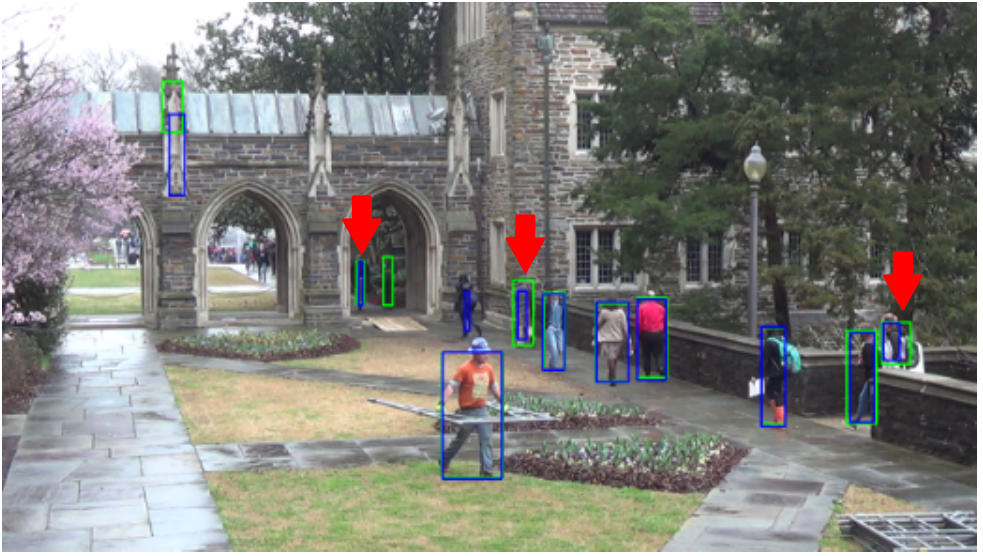}}
  \caption{The impact of accuracy reduction on video frame modification by the latency controller for JAAD (a) and DukeMTMC (b). Green and blue bounding boxes indicate pedestrian detections on unmodified and modified video frames respectively. The video frames experience an accuracy loss of $\sim$3\% for JAAD and $\sim$4\% for DukeMTMC respectively. The red arrows show the resulting detection errors.}
  \label{fig:F1_error_images}
\end{figure}

Figure \ref{fig:F1_error_images} gives a qualitative illustration of the effect of the accuracy loss with video frames from JAAD (Figure \ref{fig:F1_error_jaad}) and DukeMTMC (Figure \ref{fig:F1_error_duke}) datasets with complex scene dynamics. Pedestrian detections for the unmodified video frames (green), and after video frame modification (blue) by the network latency controller are shown (accuracy loss of 3.3\% for JAAD and 4.2\% for DukeMTMC). For the JAAD video frame, a detection error occurs at the area indicated by the red arrow. A group of two individuals are detected correctly in the unmodified video frame, but detected as a single box in the modified video frame. For the DukeMTMC frame, the detection boxes in the modified video frame are thinner than those in the unmodified video frame. We note that for both datasets, apart from the aforementioned detection errors, the other detections in the modified frame are same as the unmodified frame. 

\subsection{Mez Evaluation}
\label{sec:mez_eval}

We evaluate the pub-sub latency performance of Mez both with scaling of peer IoT camera nodes, and with scaling the number of subscriber applications. We compare Mez with the state-of-the-art low-latency NATS \cite{Nats} pub-sub messaging system. The pub-sub latency is the end-to-end time taken for a video frame to be published by the camera and subscribed by the application. The pub-sub latency includes the Publish and Subscribe API completion times, network latency, video frame modification times by controller (for Mez), and all processing delays inside the messaging system. Note that the pub-sub latency does not include the compute time for the pedestrian detection application. 

\begin{figure}
\centering
\subfloat[]{\includegraphics[width=0.48\textwidth, keepaspectratio
]{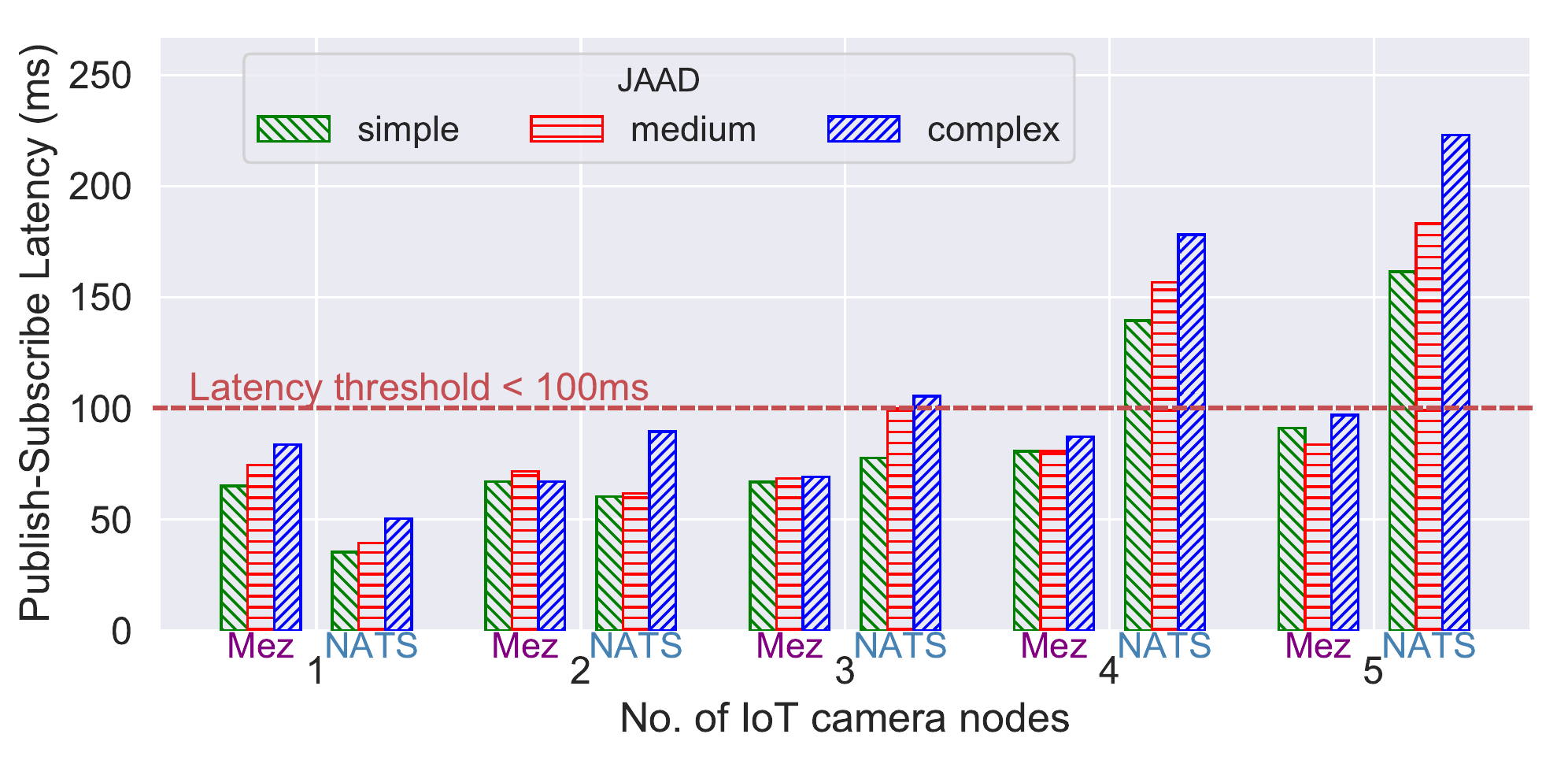}\label{fig:node_scal_jaad_latency}}\hspace{2pt}
\subfloat[]{\includegraphics[width=0.48\textwidth, keepaspectratio]{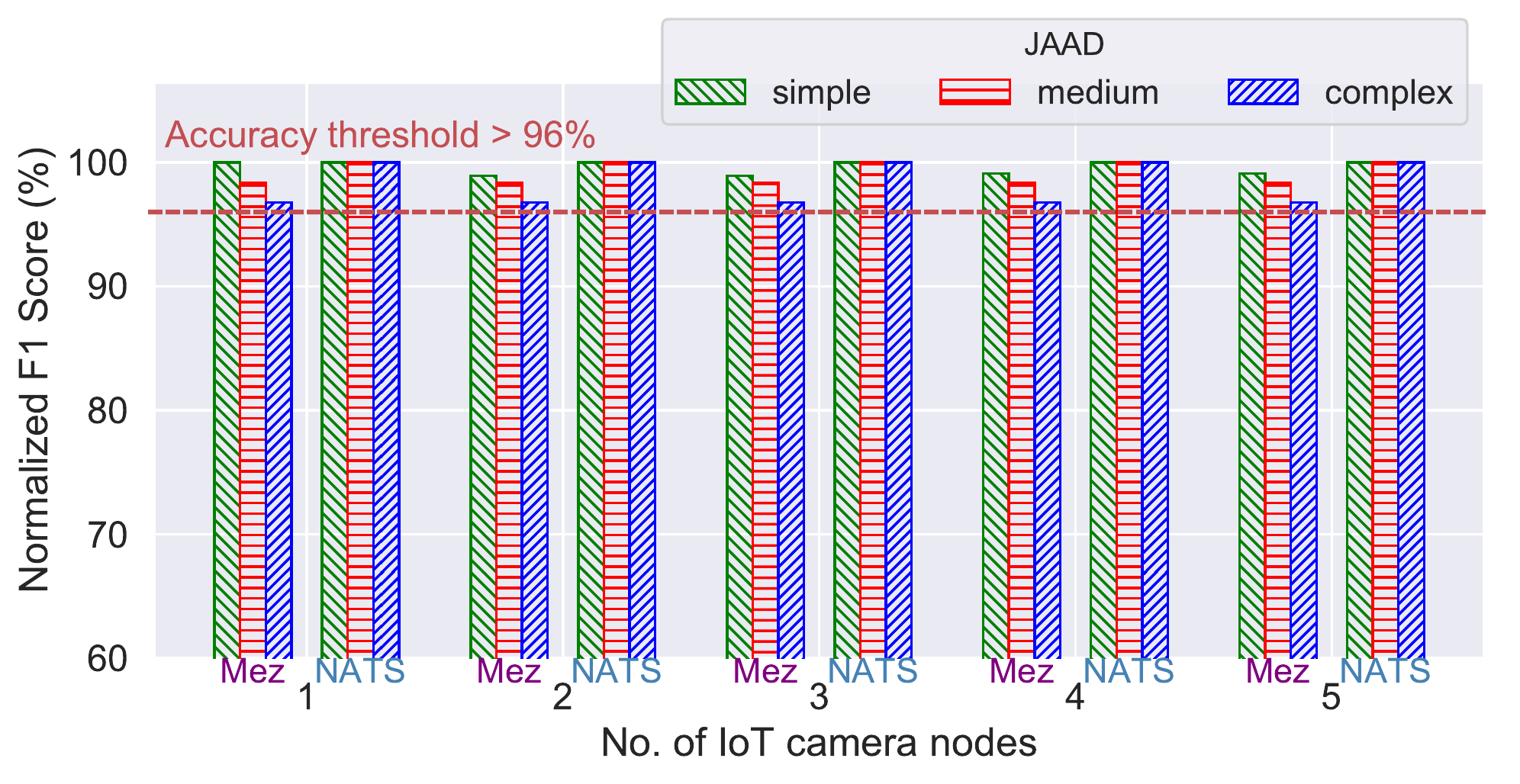}\label{fig:node_scal_jaad_acc}}
\caption{Pub-sub latency (95th percentile), and pedestrian detection accuracy with IoT node scaling for Mez and NATS for JAAD dataset with simple, medium and complex scene dynamics video frames. In (a) Y axis shows the per frame Publish-Subscribe latency. In (b) Y axis shows the accuracy in terms of the normalized F1 score percentage. X axis of both figures indicates the number of IoT camera nodes. Unlike NATS, Mez is able to achieve the latency threshold of 100ms as the number of IoT camera nodes scale. The resulting loss of accuracy is less than 3.3\%.}
\label{fig:node_scal_jaad}
\end{figure}

\begin{figure}
\centering
\subfloat[]{\includegraphics[width=0.48\textwidth, keepaspectratio
]{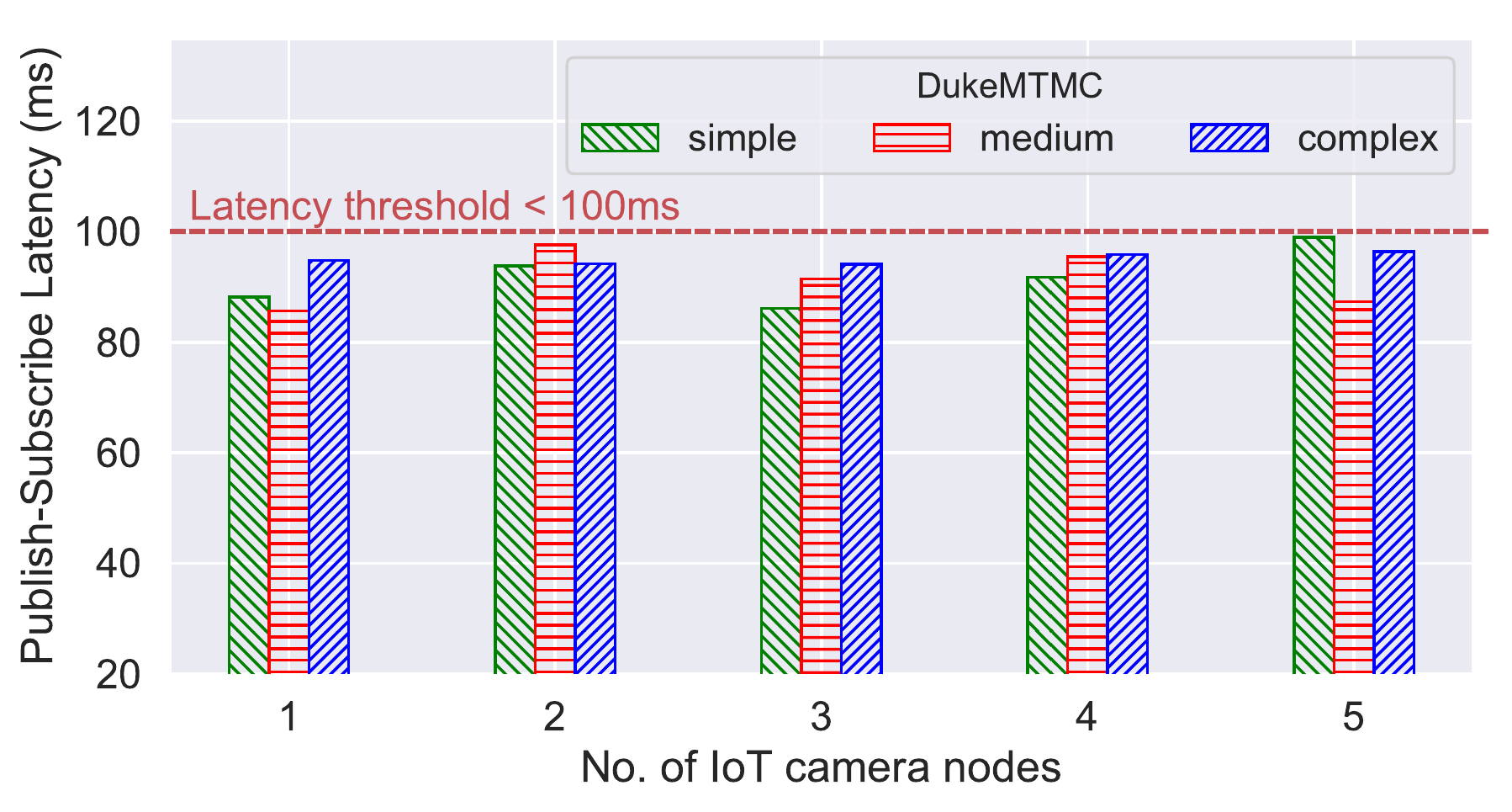}\label{fig:node_scal_duke_latency}}\hspace{2pt}
\subfloat[]{\includegraphics[width=0.48\textwidth, keepaspectratio]{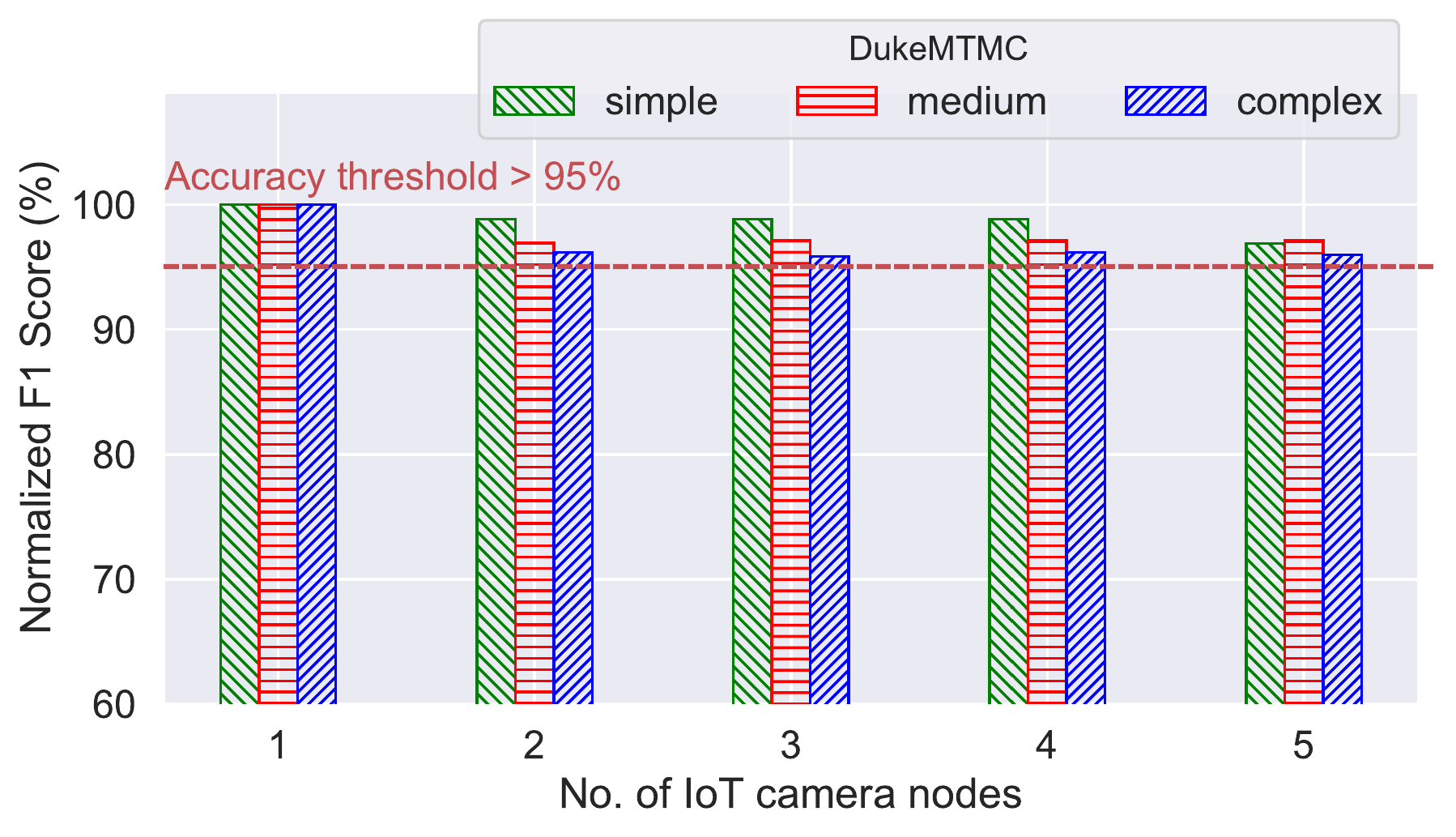}\label{fig:node_scal_duke_acc}}
\caption{Pub-sub latency (95th percentile), and pedestrian detection accuracy with IoT node scaling for Mez for DukeMTMC dataset with simple, medium and complex scene dynamics video frames. In (a) Y axis shows the per frame Publish-Subscribe latency. In (b) Y axis shows the accuracy in terms of the normalized F1 score percentage. X axis of both figures indicates the number of IoT camera nodes. Mez is able to achieve the latency threshold of 100ms as the number of IoT camera nodes scale. The resulting loss of accuracy is less than 4.2\%. Since NATS has a 1MB message size limit, DukeMTMC frames cannot be sent/received using NATS. }
\label{fig:node_scal_duke}
\end{figure}

Figure \ref{fig:node_scal_jaad_latency} shows the per frame pub-sub latency for Mez and NATS for the JAAD dataset as the number of IoT camera nodes is scaled from 1 to 5. The latency and the normalized $F1$ accuracy thresholds are set at 100 ms, and 96\% respectively. This setup emulates a scenario where a single subscriber (for example, a machine vision application for object re-identification across multiple camera views) requests video frames from multiple IoT camera nodes. As seen from Figure \ref{fig:node_scal_jaad_latency}, when the number of IoT nodes are increased, Mez is able to maintain the pub-sub latency under 100ms. However, the pub-sub latency of NATS shows a super-linear increase with IoT node scaling, since NATS does not perform any type of network latency control. Figure \ref{fig:node_scal_jaad_acc} shows the accuracy (normalized $F1$\%) achieved by Mez and NATS as the IoT camera nodes scale. Since NATS always sends unmodified images, it maintains the maximum accuracy for all cases. Mez shows a worst case accuracy reduction of 3.3\%. 

\begin{figure}
\centering
\subfloat[]{\includegraphics[width=0.75\textwidth, keepaspectratio
]{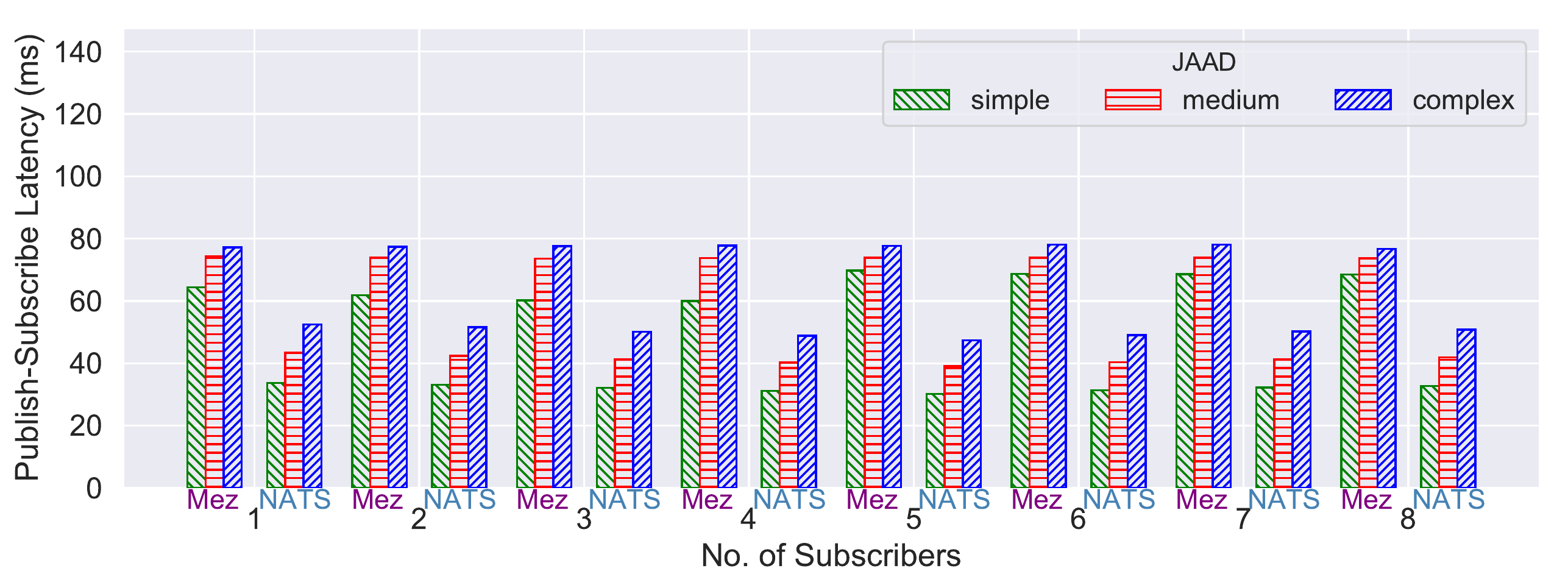}\label{fig:sub_scal_jaad}}\hspace{2pt}
\subfloat[]{\includegraphics[width=0.75\textwidth, keepaspectratio]{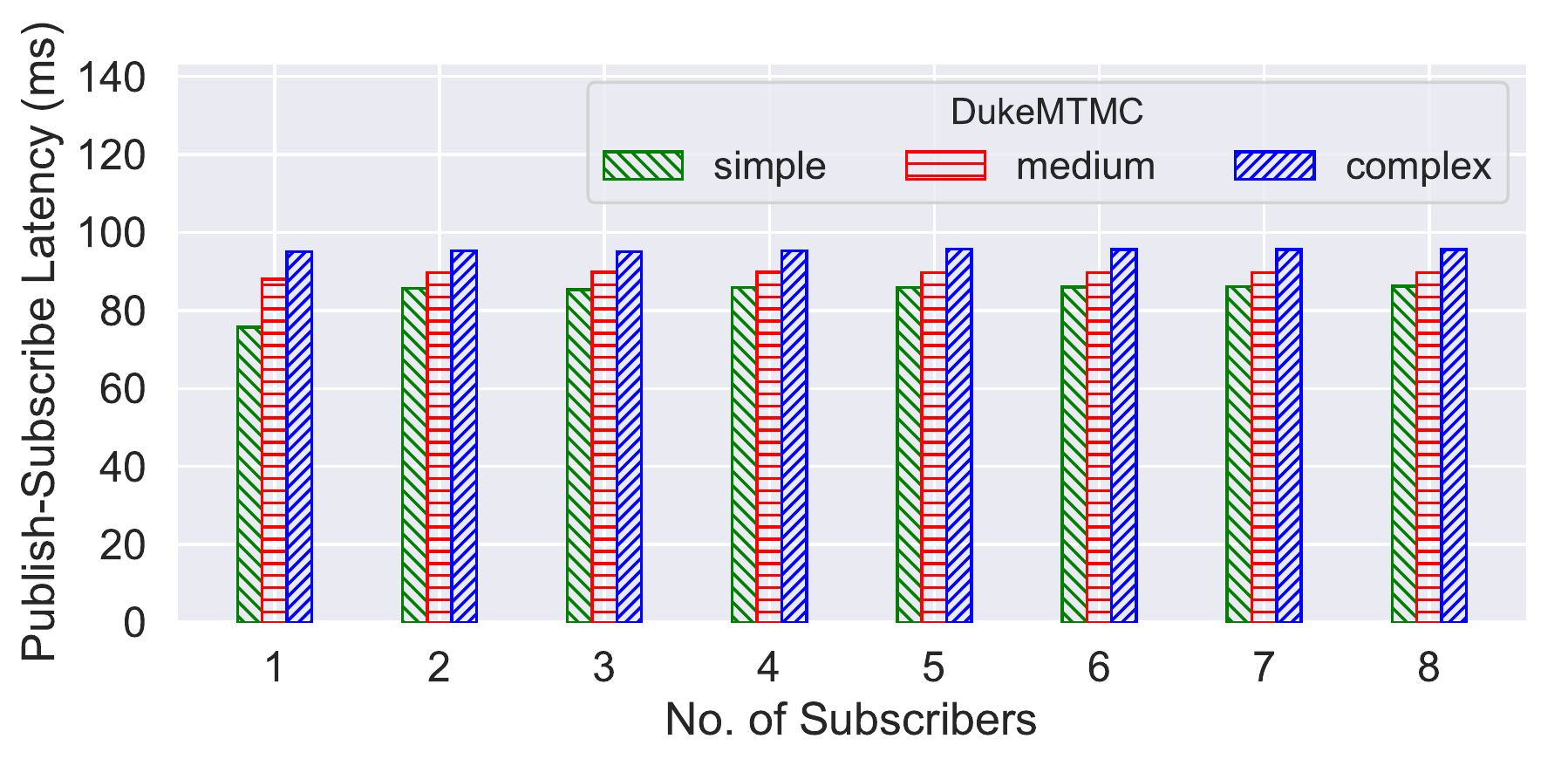}\label{fig:sub_scal_duke}}
\caption{Subscriber scaling for Mez and NATS for (a) JAAD and (b) DukeMTMC datasets with simple, medium and complex scene dynamics. Y axis shows the per frame Publish-Subscribe latency and X axis shows the number of subscribers.}
\label{fig:sub_scale}
\end{figure}

Similar evaluation is performed on the DukeMTMC dataset with the latency threshold set at 100ms for video frames with simple, medium and complex scene dynamics. Since the median video frame size for DukeMTMC is greater than 1 MB, we could not evaluate NATS due to its 1 MB message size limit.  When the number of IoT nodes is scaled from 1 to 5, Mez is able to maintain the pub-sub latency under 100ms (Figure \ref{fig:node_scal_duke_latency}) while maintaining the accuracy reduction less than 4.2\%(Figure \ref{fig:node_scal_duke_acc}).

Figure \ref{fig:sub_scale} shows the pub-sub latency for Mez and NATS as the number of subscribers are scaled. Poor subscriber scaling would indicate concurrency limitations. This set up emulates the operational scenario at the IoT-Edge where multiple vision applications (subscribers) request video frames from a single camera. In this case, since only a single IoT camera node (to which the producer is publishing video frames) is operational, there is no channel interference due to peer IoT camera nodes. Both Mez and NATS scale well as the number of subscribers are increased from 1 to 8 with minimal degradation in latency. However, Mez has a higher latency than NATS due to controller overheads.

\begin{figure}[h]
\centering
\subfloat[]{\includegraphics[width=0.4\textwidth, keepaspectratio
]{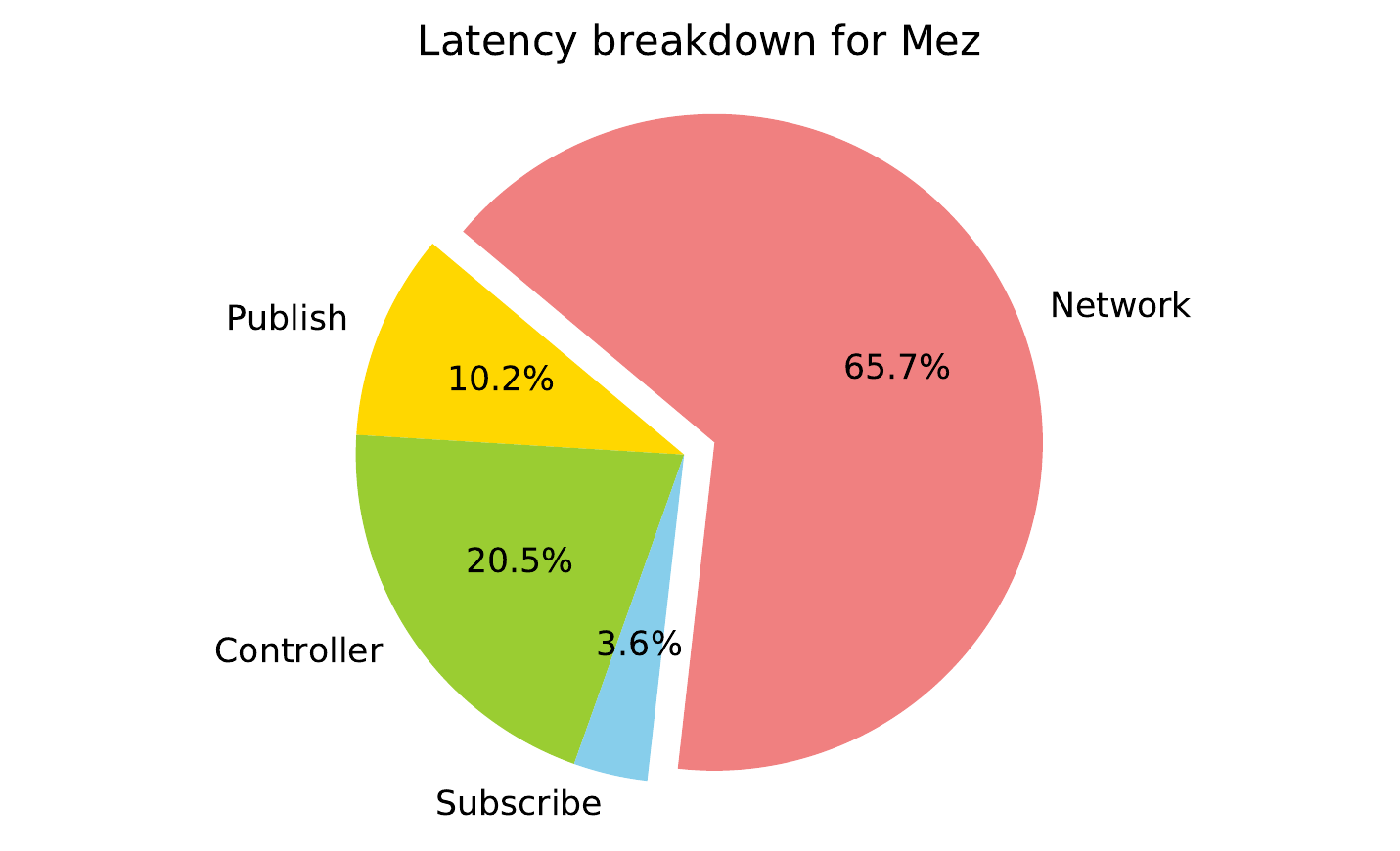}\label{fig:mez_division_with_int}}\hspace{2pt}
\subfloat[]{\includegraphics[width=0.4\textwidth, keepaspectratio]{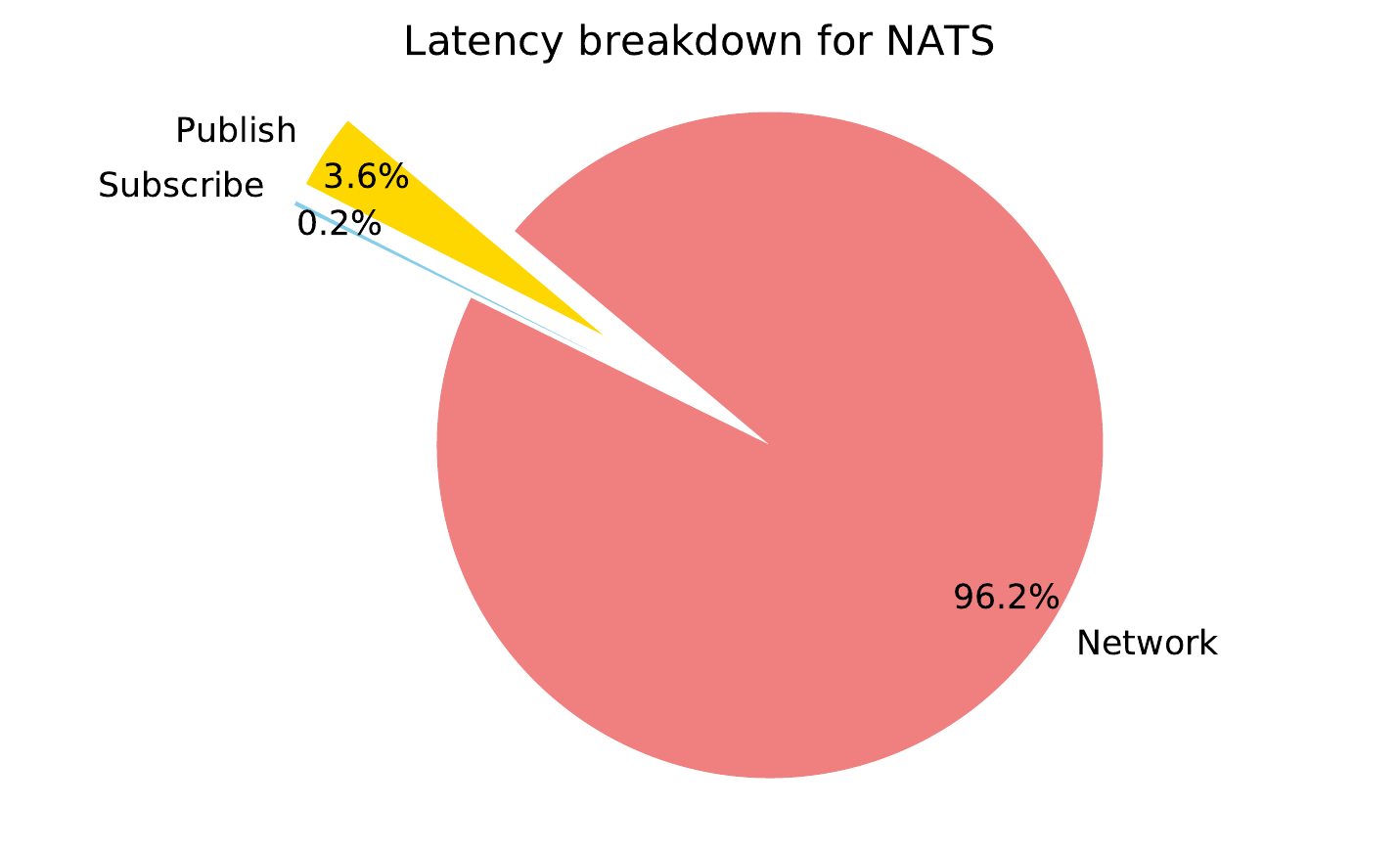}\label{fig:nats_division_with_int}}
\caption{End-to-end latency breakdown for (a) Mez and (b) NATS in presence of 4 peer nodes. The measurements are taken for complex scene dynamics video frames from the JAAD dataset with median image size of 970KB, streamed at 5fps rate. For NATS, the network latency dominates the overall latency at 96.2\%. For Mez, the network latency is the dominant component at 65.7\%, with the controller overhead being the next highest at 20.5\%  }
\label{fig:lat_breakdown}
\end{figure}

Figure \ref{fig:lat_breakdown} shows the breakdown for different components of the pub-sub latency for Mez and NATS with all 5 IoT camera nodes transferring video frames to the Edge server. The measurements are taken for complex scene dynamics video frames from the JAAD dataset. For NATS, the network latency dominates the overall latency at 96.2\%. For Mez, the network latency is the dominant component at 65.7\%, with the controller overhead being the next highest at 20.5\%. About half the controller processing time is due to the video frame modification, with the video frame copying between the logs at IoT camera nodes accounting for the remaining time. We are currently investigating the use of GPUs available on the Nvidia Xavier boards to lower the video frame processing time. Integrating the controller as a part of the CamBroker instead of the current approach of the controller as a separate microservice, could result in lowering the video frame copying overheads.

\section{Discussion and Future work}
\label{sec:discussion}

In this section we review the different choices made in the design of Mez, and discuss alternatives. 

\textbf{GPU computing at IoT nodes: }
In our work we have only used the 8 core ARM CPU available on the Nvidia Xavier board. By using the GPU available on the Nvidia Xavier board, additional computationally intensive video frame modifications including performing object detection at the IoT camera node could be employed. 

\textbf{Video frame compression: }
 The video frames that are transferred from the IoT camera node to the Edge server could be compressed (for example, using H.264) to reduce the data size. However, Canel et. al. \cite{scaling_video} suggest that a low quality H.264-encoded 1080p (1920x1080 pixels) stream is insufficient to perform accurate analysis for vision analytic applications such as traffic monitoring and pedestrian tracking.

\textbf{Availability: }
Regarding service availability, the current version of Mez does not support replicated physical nodes (neither IoT camera nodes, nor Edge sever). For hardware failures, such as if an IoT node fails, the video frames from the associated camera becomes unavailable. However, if the Edge server fails, then entire system becomes unavailable. The ability to support replicated Edge server components (logs and the EdgeBroker) is part of future work. However, we note that computational resources are constrained at the Edge due to power, space, and cost considerations. Thus, physical replication of resources may not be practical on the Edge. An alternative worth investigating is the use of Cloud for fail over of the Edge server so that the system is still operational, albeit at a reduced performance.

For software failures, Mez could rely on container orchestrators such as Kubernetes \cite{kubernetes} for resurrecting failed services. In this case, Mez needs to be containerized (for example, using Docker containers \cite{docker}), with Kubernetes orchestrating the containers. The microservice architecture of Mez allows ready containerization of the brokers, and the latency controller. A Continuous Integration/Continuous Delivery framework (for example, with Jenkins \cite{jenkins}) could also be used to ease the deployment and updating of Mez without loss of service. Note that Kubernetes itself will need hardware redundancy to guarantee its availability. With limited hardware, an interesting possibility is to use the Cloud to ensure availability of Kubernetes. 

\textbf{Security: }
 Mez takes advantage of the security features of gRPC to implement TLS based certificate authentication for clients and servers, as well as TLS based encryption of video frames in transit. Additionally, all sensitive video frames at rest that are persisted on disk are encrypted. We leave the implementation of additional security measures such as Role Based Access Control (RBAC) for application access to video frame data, and the verification of the authenticity of container images to future work. 
 
\begin{figure}
  \begin{center}
    \includegraphics[width=0.5\textwidth]{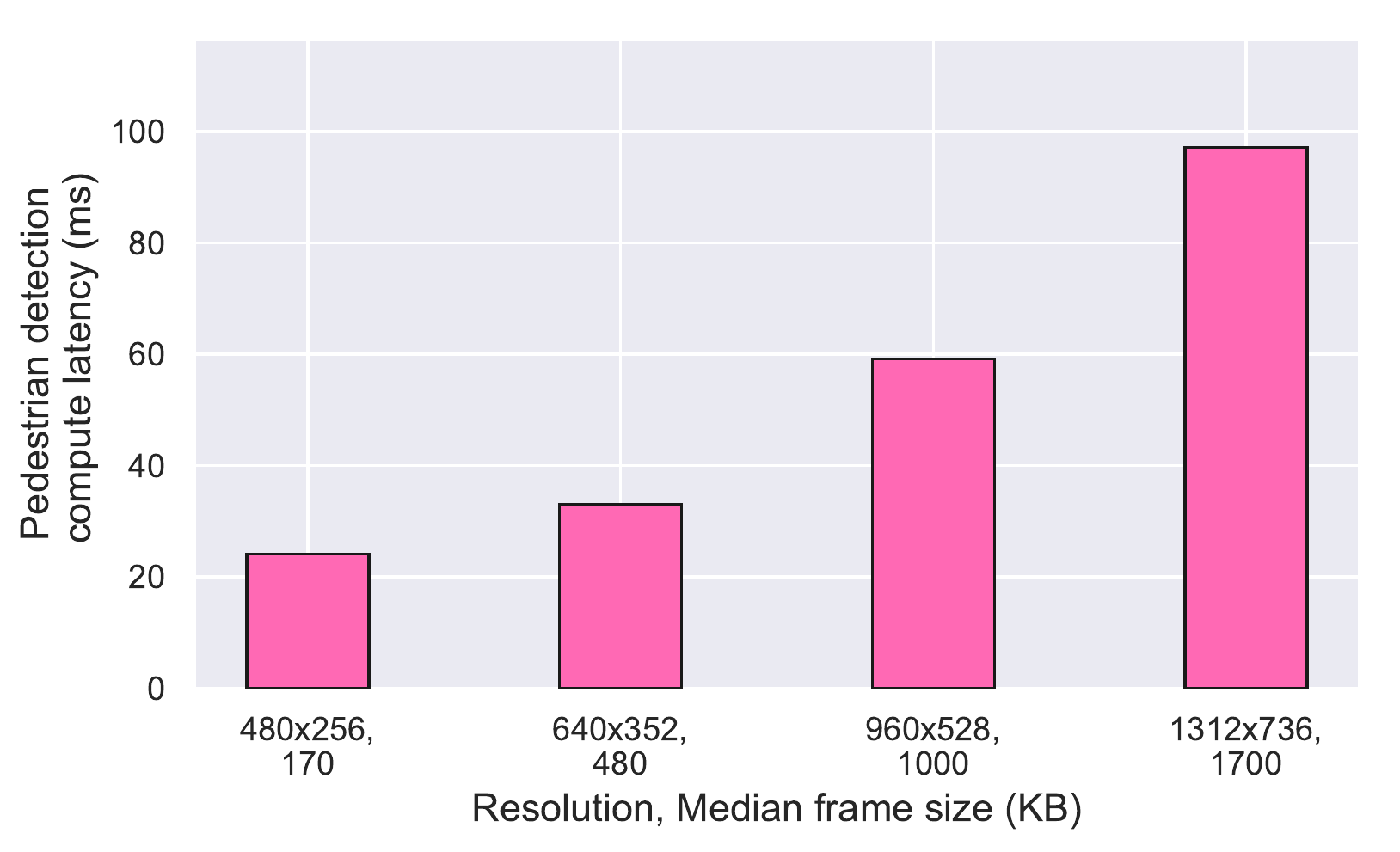}
  \end{center}
  \caption[Images from JAAD and DukeMTMC datsets.]{\label{fig:obj_det_latency}Compute latency for pedestrian detection with OpenPose (on Nvidia Titan V GPU) vs. video frame size }
\end{figure}

\textbf{Compute latency: }
In Figure \ref{fig:obj_det_latency} we investigate the dependence of the compute latency on the video frame size. We note that, the compute latency is only dependent on the resolution of the images. Since one of the tuning knobs is image resolution, with reduced resolution and with reduced image size, OpenPose could achieve reduction in compute latency as well.

Figure \ref{fig:obj_det_latency} shows that the pedestrian detection compute latency increases with increase in resolution and image size.  The measurements are done on an Nvidia Titan V GPU. In this paper we have only focused on the network latency. However, due to the compute intensive nature of these applications, there exists a possibility of jointly optimizing compute and network latency, so as to obtain the overall desired latency. We leave the exploration of this topic to future work. 

\section{Related work}
\label{sec:related_work}

In this Section, we describe previously reported work on Edge computing, application of approximate computing, and distributed messaging systems.

\subsection{Edge Computing}

The concept and motivation behind Edge computing are described in a number of recent publications \cite{case_for_VM, swarm_at_edge_cloud, edge_comp_challenge, Fog-edge, Fog-computing, fog-princeton, IEEE-mag-MEC, MEC-city, IoT-app}. Regarding machine vision at the Edge, in the Gabriel project \cite{Ha2014Gabriel}, Ha et. al. describe a wearable cognitive assistance system where the images captured by a mobile device are processed by the Edge node to analyze what the user is seeing, and provide the user with cues as to what is in the scene (for example, recognizing a person). In the VisFlow project, Lu et al. \cite{Lu2016visflow} describe a system that can analyze feeds from multiple cameras for license plate recognition and real-time traffic flow mapping. In \cite{revamp2t}, Neff et. al. proposes REVAMP$^{2}$T, an IoT system that tracks pedestrians across multiple cameras by running custom-designed deep learning based vision engines at the low power Edge nodes close to the cameras. However, none of these works address guaranteeing of latency requirements at the Edge for machine vision applications. 

In the Hetero-Edge project, Zhang et. al. \cite{Zhang2019} describe a system that can efficiently orchestrate real-time vision applications on heterogeneous Edge servers. The new resource orchestration platform developed, uses a set of task scheduling schemes to make the Hetero-Edge system latency-aware, but does not consider communication latency. In \cite{rein_video}, Pakha et. al. introduce the idea of control knobs such as frame selection and area cropping to parametrize a custom video protocol that streams videos from cameras to Cloud servers to perform neural-network-based video analytics. Ther work highlights opportunities to improve the tradeoffs between bandwidth usage and inference accuracy, but does not address Edge specific latency requirements demanded by many IoT vision applications.

\subsection{Approximate Computing}
 
In \cite{approx_comp_mittal}, Mittal provides a survey of approximate computing techniques. Strategies for approximation at the code level such as loop perforation, and at the architecture level such as reduced precision operations are discussed.  Regarding applications of approximate computing to Deep Learning, Chen et al. \cite{Chen18}  use approximate computing to accelerate network training, while Ibrahim et. al. \cite{ibrahim2018approximate} explore the use of approximate computing to realize Deep Learning networks on resource constrained embedded platforms. Unlike our work, in these works approximate computing is targeted towards reducing the computational load.  

In \cite{betzel2018approximate}, Betzel et. al. introduce the concept of approximate communication to reduce the communication between processing elements in a high performance computing system. They evaluate compression, reduced synchronization, and value prediction as potential approximate communication techniques. In contrast to Betzel et. al. we target latency variations due to interference in wireless communication channels, and investigate the impact on application accuracy.

\subsection{Distributed Messaging Frameworks}
RabbitMQ \cite{RabbitMQ} is an open source messaging system that supports the Advanced Message Queuing Protocol (AMQP), Streaming Text Oriented Messaging Protocol (STOMP), Message Queuing Telemetry Transport (MQTT), and other protocols. It supports multiple messaging styles including pub-sub, request-reply, and point-to-point communication models. RabbitMQ's design assumes a smart broker, dumb consumer model, with the broker consistently delivering messages. Mez, in contrast, supports a dumb broker, smart consumer model, and is designed specifically for machine vision applications at the Edge.

Kafka proposed by Kreps et al. in \cite{KrepsKafka} is used for collecting and delivering high volumes of data with high throughput. It combines the benefits of traditional log aggregators and messaging systems. Kafka is a pub-sub system in which multiple producers and consumers can publish and retrieve messages at the same time, and store streams of data in distributed, fault tolerant clusters using multiple brokers and partitions.
Similar to Mez, Kafka supports a dumb broker, smart consumer model. However, Kafka is focused on delivering high throughput, and not necessarily on latency of individual messages. 
 
NATS \cite{Nats} messaging system is a recent project that is focused on providing low latency to cloud native applications. Similar to Mez, NATS supports a pub-sub system, and a dumb broker, smart consumer model. However, unlike Mez, NATS is a general purpose messaging system, and does not provide latency guarantees.


\section{Conclusions}
\label{sec:conclusion}

In this paper, we propose Mez - a publisher-subscriber messaging system for distributed machine vision at the IoT Edge. Mez provides machine vision applications at the Edge the ability to specify network latency upper bound for the video frames transferred from the IoT camera nodes to the Edge server, along with an accuracy lower bound that the application can tolerate. Many machine vision applications support the approximate computing paradigm, where useful enough results can be obtained despite reduced quality input. The latency controller in Mez achieves application specified network latency despite channel interference, by modifying the video frames to reduce their sizes such that the application accuracy specifications are satisfied as well. Additionally, the design of Mez incorporates an in-memory log based storage that takes advantage of specific features of machine vision applications to implement low latency operations. We also discuss the fault tolerance capabilities of the Mez design. Our experimental evaluation of Mez on an IoT Edge testbed with a pedestrian detection machine vision application indicates that Mez is able to tolerate latency variations of up to 10x with a modest drop in application accuracy of less than 4.2\%. We also discuss alternatives to some of our design choices, and suggest directions for future work.


\begin{acks}
This research is partially supported by the National Science Foundation (NSF) under Award No. 1831795.
\end{acks}

\bibliographystyle{ACM-Reference-Format}
\bibliography{REFERENCES}

\appendix
\end{document}